\documentclass[11pt,a4paper]{article}
\pdfoutput=1
\usepackage[pdftex]{graphics}
\usepackage{jheppub}
\usepackage{amsmath,amssymb,amsfonts}
\usepackage{enumitem}
\usepackage{multirow}
\usepackage{array,booktabs}
\usepackage{tikz}

\newcommand{\be}{\begin{eqnarray}}
\newcommand{\ee}{\end{eqnarray}}
\newcommand{\nn}{\nonumber}
\newcommand{\bn}{\begin{enumerate}}
\newcommand{\en}{\end{enumerate}}
\newcommand{\bl}{\begin{align}}
\newcommand{\el}{\end{align}}

\def\thalf{{\textstyle \frac{1}{2}}}

\def\det{{\rm det}}

\def\jmath{{j}}

\title{ABJM Amplitudes in U-gauge and a Soft Theorem}

\author[a]{Seungbeom Chin,} 
\author[a,b,c,d]{Sangmin Lee,} 
\author[a]{Youngbin Yun\,} 

\affiliation[a]{Department of Physics and Astronomy, Seoul National University, Seoul 151-747, Korea}
\affiliation[b]{Center for Theoretical Physics, Seoul National University, Seoul 151-747, Korea}
\affiliation[c]{College of Liberal Studies, Seoul National University, Seoul 151-742, Korea}
\affiliation[d]{School of Physics, Korea Institute for Advanced Study, Seoul 130-722, Korea}

\abstract{
We report progress in computing and analyzing all tree amplitudes in ABJM theory. Inspired by the isomorphism between the orthogonal Grassmannian and 
the pure spinor geometries, we adopt a new gauge, called u-gauge, for evaluating the orthogonal Grassmannian integral for ABJM amplitudes. We carry out the integral explicitly for the 8-point amplitude and obtain the complete supersymmetric amplitude. 
The physical and spurious poles arise from the integral as expected from on-shell diagrams. 
We also derive a double soft theorem of ABJM amplitudes and verify it 
for known amplitudes.}

\arxivnumber{1508.07975}

\preprint{SNUTP15-008, KIAS-P15043}

\begin{document}

\maketitle

\section{Introduction and discussions}

Recent years have witnessed major breakthroughs in computing and understanding scattering amplitudes of gauge theory and gravity (see, e.g.,  \cite{Elvang:2013cua} for reviews). At many stages of the development, the $\mathcal{N}=4$ supersymmetric Yang-Mills theory (SYM) proved to be an extremely fruitful testing ground. Many novel ideas, such as twistor string theory \cite{Witten:2003nn}, dual superconformal symmetry \cite{Drummond:2008vq}, 
Grassmannian formulation \cite{ArkaniHamed:2009dn}, on-shell diagram representation \cite{ArkaniHamed:2012nw} and amplituhedron \cite{Arkani-Hamed:2013jha,Arkani-Hamed:2013kca}, are realized in their simplest forms in $\mathcal{N}=4$ SYM and then generalized to less symmetric theories. 

The three dimensional $\mathcal{N}=6$ supersymmetric Chern-Simons-matter theory \cite{Aharony:2008ug,Hosomichi:2008jd,Benna:2008zy,Hosomichi:2008jb,Bagger:2008se,Bandres:2008ry,Schnabl:2008wj}, often referred to as the ABJM theory, is a close sibling of the $\mathcal{N}=4$ SYM in many respects. For instance, the scattering amplitudes of the ABJM theory exhibit dual superconformal symmetry \cite{Bargheer:2010hn,Huang:2010rn,Huang:2010qy,Gang:2010gy} and admits Grassmannian \cite{Lee:2010du}, twistor string \cite{Huang:2012vt,Cachazo:2013iaa,Engelund:2014sqa}, and on-shell diagram formulations \cite{Huang:2013owa,Kim:2014hva,Huang:2014xza}. Despite these parallel successes, the study of ABJM amplitudes fall short of those of $\mathcal{N}=4$ SYM in many respects. One of the most pressing problem is the lack of a ``momentum twistor" formulation \cite{Hodges:2009hk,Mason:2009qx,ArkaniHamed:2009vw} in which the dual superconformal symmetry would become manifest (see \cite{Elvang:2014fja} for a recent attempt in this direction).   
Closely related to momentum twistors are the dual superconformal $R$-invariants \cite{Drummond:2008vq}, which serves as building blocks 
for an explicit formula for all tree amplitudes \cite{Drummond:2008cr} and a starting point for the construction of amplituhedron. 

In principle, all ingredients to compute the ABJM tree amplitudes are available in the literature. The Grassmannian integral \cite{Lee:2010du}, supplemented by the contour prescription from on-shell diagrams \cite{Huang:2013owa,Kim:2014hva,Huang:2014xza}, will produce the amplitudes. A mundane, yet seemingly unavoidable, problem is that each BCFW bridge in the on-shell diagram introduces a quadratic equation in the integration variables. 
The solutions to quadratic equations generically contain square-roots, 
which must cancel out when summed over all solutions and produce a rational function of kinematic variables. Mainly for this technical reason, explicit results for ABJM tree amplitudes to date are limited to 4- and 6-point amplitudes \cite{Bargheer:2010hn,Lee:2010du,Gang:2010gy} which are free from square-roots due to limited kinematics, and a partial result for 8-point amplitude \cite{Gang:2010gy} without manifest supersymmetry. 

The goal of this paper is to take a few steps toward the computation of all ABJM tree amplitudes. Our two main results are a complete evaluation of the supersymmetric 8-point amplitude and a derivation of a double soft theorem valid for all tree amplitudes.

In evaluating the Grassmannian integral for 8-point or higher amplitudes, 
we find it convenient to use a new gauge, which we call ``u-gauge". The u-gauge is inspired by the isomorphism between the orthogonal Grassmannian and 
the pure spinor geometries; both of them admit the $SO(2k)/U(k)$ coset description. A particular set of coordinates of the coset space introduced in \cite{Berkovits:2004bw} trivially solves the orthogonality constraint and can be easily generalized to arbitrary $k$. This fact makes the u-gauge, at least in some contexts, more convenient than conventional gauges involving Euler angle coordinates. 

Although the u-gauge do not circumvent the square-root problem mentioned above, the quadratic equations in the u-gauge tend to be simpler, which allow us to combine all residues in the contour integral. For the 8-point amplitude, 
the integral is effectively one-dimensional. 
We can express the denominators of the amplitude in terms of the standard cross-ratios among solutions to quadratic equations. It is easy to see that 
the cross-ratios can in turn written in terms of the coefficients of the quadratic equations, thereby avoiding the need to solve the equations explicitly.

The final result for the 8-point amplitude takes the form, 
\begin{align}
\mathcal{A}_{8} = \delta^3(P) \delta^6(Q)   (1+\pi)  J_8^B J_8^F \left(
\frac{F(1)}{\Delta_{12}\Delta_{13}\Delta_{14}} +\frac{F(3)}{\Delta_{31}\Delta_{32}\Delta_{34}}
\right)  \,.
\end{align}
The $(1+\pi)$ factor accounts for the sum over two disjoint branches of the orthogonal Grassmannian. The two rational functions in the big parenthesis corresponds to the two on-shell diagrams contributing to the 8-point amplitude.
The numerators $F(1)$, $F(3)$ as well as the $\Delta_{ij}$ factors in the denominators are polynomials in kinematic variables. 
The on-shell diagrams suggest that 
$\Delta_{12}, \Delta_{14}, \Delta_{32}, \Delta_{34}$ should be proportional to physical poles of the amplitude whereas $\Delta_{13} = \Delta_{31}$ should be  spurious. We confirm the expectation by explicitly proving that $\Delta_{ij}$ for the physical poles are proportional to $p_{klm}^2$ factors for adjacent particles. 

In the second half of this paper, we consider the double-soft limit of ABJM amplitudes. Soft limits of scattering amplitudes in gauge and gravity are well known to exhibit universal behavior and have bearing on gauge symmetries and spontaneously broken global symmetries. Our motivation to study the soft limit is more modest. As we make progress in computing higher point amplitudes, 
we wish to use the soft theorem to test the consistency of the methods we use. 
Our derivation of the soft theorem will closely follow that of ref.~\cite{Chen:2014cuc}, where a similar double-soft theorem was derived for three dimensional supergravity theories. 
We show that the $(2k+2)$-point amplitude $\mathcal{A}_{2k+2}$ reduces to the $2k$-point amplitudes $\mathcal{A}_{2k}$ with universal leading and sub-leading soft factors in the double soft limit,
\begin{equation}
\left. \mathcal{A}_{2k+2}(1, 2, \cdots,2k, \epsilon^2 p_{2k+1},\epsilon^2 p_{2k+2}) \right|_{\epsilon \rightarrow 0}
= \left( \frac{1}{\epsilon^2} S^{(0)} + \frac{1}{\epsilon} S^{(1)} \right) \mathcal{A}_{2k}(1,2,\cdots,2k)\,.
\end{equation}
As in \cite{Chen:2014cuc}, the proof of the soft theorem is based on the BCFW \cite{Britto:2004ap,Britto:2005fq} recursion relation of the ABJM theory \cite{Gang:2010gy}.
We confirm that the universal soft factors respect all the symmetries of 
the ABJM amplitudes for all $k$. For 6-point amplitude to the sub-leading order, 
and for 8-point amplitude to the leading order, we explicitly take the soft limit of the known amplitude and verify that the soft theorem holds. 

Although we still have explicit form of tree amplitudes only up to 8-point, 
we expect that the findings in the present paper, such as the u-gauge, cross-ratios among different poles in the contour integral, and the double soft theorem, will lay the groundwork for a complete construction of all tree amplitudes of ABJM theory in terms of momentum twistors and/or dual superconformal $R$-invariants.

This paper is organized as follows. 
In section 2, we give a short review of the general structure of the ABJM tree amplitudes and the Grassmannian integral. Then we introduce the u-gauge and compare it with other well-known gauges.
In section 3, we use the u-gauge to compute some tree amplitudes. After reproducing the 4-point and 6-point amplitudes, we present the details of how to evaluate the 8-point amplitude. 
In section 4, we propose the double soft theorem of the ABJM amplitudes and prove it using the BCFW recursion relation. We take the double soft limit of the 6- and 8-point amplitudes, and verify explicitly that the theorem holds. 

\section{Grasssmannian integral in the U-gauge}

\subsection{Preliminaries}

Here we briefly review the 3-dimensional spinor helicity formalism \cite{Bargheer:2010hn} and introduce the Grassmannian integral formula for tree level amplitudes of planar ABJM theory \cite{Lee:2010du}.

Each spinor in three dimensions transforms under SL(2, $\mathbb{R})$, and a null momentum can be written in the bi-spinor form 
\begin{align}
p^{\alpha\beta}=p_\mu(\sigma^\mu)^{\alpha\beta}=\lambda^\alpha \lambda^\beta.
\label{pll}
\end{align}
Our convention for spinors and gamma matrices are such that $p^{\alpha\beta}$ is real for real $p_\mu$, and $\lambda^\alpha$ is real (purely imaginary) for outgoing (incoming) particles. The spinors are contracted as $\langle ij \rangle \equiv \lambda_i^{\alpha}\lambda_{j\alpha}$. We normalize the norm of  vectors such that
$p_{ij}^2 = (p_i + p_j)^2 = \langle ij\rangle^2$ when both 
$\lambda_i$ and $\lambda_j$ are real.  

The on-shell superfield notation for ABJM amplitudes is built on three fermionic coordinates $\eta^I$, in addition to $\lambda^\alpha$, which transform as $\mathbf{3}$ under the U(3) subgroup of the SO(6) R-symmetry group. The particle/anti-particle superfields take the form
\begin{align}
\Phi=\phi^4 +\eta^I\psi_I +\frac{1}{2}\epsilon_{IJK}\eta^I\eta^J\phi^K+\frac{1}{6}\epsilon_{IJK}\eta^I\eta^J\eta^K\psi_4, \\
\bar{\Phi}=\bar{\psi}^4 +\eta^I\bar{\phi}_I +\frac{1}{2}\epsilon_{IJK}\eta^I\eta^J\bar{\psi}^K+\frac{1}{6}\epsilon_{IJK}\eta^I\eta^J\eta^K\bar{\phi}_4.
\end{align}
A collective notation $\Lambda =(\lambda;\eta)$ will be used when appropriate.
The fact that \eqref{pll} is invariant under $\lambda \rightarrow -\lambda$, while the wave-functions of fermions pick up a minus sign, implies 
the so-called ``$\lambda$-parity" of the super-amplitudes.  
\begin{align}
\label{coloram}
\mathcal{A}(\Lambda_1, ...,-\Lambda_i, ...,\Lambda_{2k})=(-1)^i \mathcal{A}(\Lambda_1, ...,\Lambda_i,...,\Lambda_{2k}).
\end{align}

The super-conformal generators of the superconformal symmetry come in three types:
\begin{align}
\Lambda\frac{\partial}{\partial\Lambda}, \qquad \Lambda\Lambda, \qquad \frac{\partial^2}{\partial\Lambda\partial\Lambda},
\end{align}
For the second type, we will use the notation
\begin{align}
p^{\alpha\beta}=\lambda^\alpha\lambda^\beta, \qquad q^{\alpha I}=\lambda^\alpha\eta^I,\qquad r^{IJ}=\eta^I\eta^J.
\end{align}
The super-momentum conservation is denoted as delta functions by
\begin{align}
\delta^3(P)\delta^6(Q) \qquad \textrm{with}\qquad P\equiv \sum_{i=1}^{2k} p_i^{\alpha\beta},\qquad Q\equiv \sum_{i=1}^{2k} q_i^{\alpha I} .
\end{align}

The Grassmannian integral formula for the tree level amplitudes of planar ABJM theory, first proposed in \cite{Lee:2010du}, is  

\begin{align}
\label{grassmannian}
\mathcal{L}_{2k}(\Lambda)=\int \frac{d^{k\times 2k}C}{\textrm{vol[GL(k)]}}\frac{\delta^{k(k+1)/2}(C\cdot C^T)\delta^{2k|3k}(C\cdot\Lambda) }{M_1M_2\cdots M_{k-1}M_k}.
\end{align}
The integration variable $C$ is a $(k\times 2k)$ matrix. The dot products denote $(C\cdot C^T)_{mn}=C_{mi}C_{ni}$, $(C\cdot \Lambda)_m=C_{mi}\Lambda_i$. The consecutive minor $M_i$ is defined by
\begin{align}
M_i=\epsilon^{m_1\cdots m_k}C_{m_1(i)}C_{m_2(i+1)}\cdots C_{m_k(i+k-1)}.
\end{align}
It was shown in \cite{Lee:2010du} that this formula satisfies the same cyclic symmetry and superconformal symmetry as the tree-level (2k)-point amplitude. 
Yangian invariance of the formula was first argued in \cite{Lee:2010du} and 
explicitly proved later in \cite{Bargheer:2014mxa}.

The integral \eqref{grassmannian} should be considered as a contour integral on the moduli space of rank $k$, $(k\times 2k)$ matrices $C$ with the constraint $C\cdot C^T=0$  and the equivalence relation $C\sim gC$ ($g\in$ GL$(k)$). This moduli space is known as the orthogonal Grassmannian OG($k,2k$). The dimension of OG($k,2k$) is determined by the aforementioned  two conditions:
\begin{align}
\textrm{dim}_{\mathbb{C}}\textrm{[OG}(k,2k)] =2k^2-k^2-\frac{k(k+1)}{2} =\frac{k(k-1)}{2}.
\end{align}

Integrating out the bosonic delta function $\delta^{2k}(C\cdot \lambda)$ 
leaves the momentum conserving delta function and a contour integral 
over $(k-2)(k-3)/2$ variables. The geometry and combinatorics behind 
the Grassmannian integral for all tree amplitudes, as well as some loop amplitudes, have been elucidated in \cite{Huang:2013owa,Kim:2014hva,Huang:2014xza}. On the other hand, explicit computation of amplitudes has never proceeded beyond 8-point \cite{Gang:2010gy}.

\subsection{U-gauge \label{sec:u-gauge}} 

We take a real slice of the complex orthogonal Grassmannian with the split signature, where the ``metric" in the particle basis 
is 
\begin{align}
g = \mathrm{diag}(-,+,-,+,\cdots)\,.
\label{split-sign}
\end{align} 
In this basis, the momenta and their spinor variables are related by 
\begin{align}
p_i^{\alpha\beta} = (-1)^i \lambda_i^\alpha \lambda_i^\beta\,, 
\quad \lambda_i^\alpha \in \mathbb{R}\,, 
\quad (p_i+p_j)^2 = (-1)^{i+j} \langle ij \rangle^2 \,.
\end{align}
We find it convenient to switch between the particle basis 
and the light-cone basis:
\begin{align}
&ds^2 = \sum_{i=1}^{2k} (-1)^i (dx_i)^2 
= \sum_{m=1}^{k} 
dw_m dv^m 
\quad
(w_m = x_{2m} + x_{2m-1}, \;v^m = x_{2m} - x_{2m-1}) \,.
\label{split-sign-2}
\end{align}
In the light-cone basis, a $GL(k)_R$ 
subgroup of the $O(k,k)$ symmetry group remains manifest.  
We will use a notation with covariance under $SL(k) \subset GL(k)_R$ and adopt the summation convention. 
The invariant tensor 
of $SL(k,\mathbb{R})$ will be denoted by $\epsilon_{m_1\cdots m_k}$. 

The spinor-helicity variable for particles are denoted by $\lambda_i^\alpha$ ($i=1,\ldots,2k$, $\alpha=1,2$). We use the same letters 
$w$, $v$ for the light-cone combinations of the spinor variables:
\begin{align}
w_m^\alpha = \lambda_{2m}^\alpha + \lambda_{2m-1}^\alpha\,, \quad 
v^{m\alpha} = \lambda_{2m}^\alpha - \lambda_{2m-1}^\alpha\,.
\end{align}
The scalar product of two spinors are defined in a usual manner.
\begin{align}
\langle \lambda_1 \lambda_2  \rangle 
= \epsilon_{\alpha\beta} \lambda_{1}^\alpha \lambda_{2}^\beta \,.
\end{align}
To avoid confusion, we reserve the shorthand notation $\langle 1 2  \rangle = \langle \lambda_1 \lambda_2  \rangle$ exclusively for the particle basis. 
In the light-cone basis, we will use $\langle w_m w_n \rangle$, 
$\langle w_m v^n \rangle$ and so on. 
The overall momentum conservation is written as 
\begin{align}
P^{\alpha\beta} 
= \sum_{i=1}^{2k} (-1)^i \lambda_i^\alpha \lambda_i^ \beta = \frac{1}{2}
\left( w_m^\alpha v^{m\beta} + w_m^\beta v^{m\alpha} \right)
=0 \,.
\label{p-cons}
\end{align} 
The light-cone components of the fermionic coordinates $\eta^I_i$ are 
denoted by 
\begin{align}
\bar{\theta}_m^I = \eta_{2m}^I +\eta_{2m-1}^I \,, \quad
\theta^{mI} = \eta_{2m}^I -\eta_{2m-1}^I \,.
\end{align}
The supermomentum components are rewritten as 
\begin{align}
Q^{\alpha I} = \sum_{i=1}^{2k} (-1)^i \lambda_i^\alpha \eta_i^I 
 =  \frac{1}{2}
\left( w_m^\alpha \theta^{mI} +  v^{m\alpha} \bar{\theta}^I_m \right) \,.
\end{align}
In summary, the metric and the kinetic variables in the light-cone basis take the form
\begin{align}
g = 
\frac{1}{2} 
\begin{pmatrix}
0 & \delta^m{}_n  \\
\delta_m{}^n & 0
\end{pmatrix} \,, 
\quad 
\lambda = 
\begin{pmatrix}
w_n \\ v^n 
\end{pmatrix} \,,
\quad 
\eta = 
\begin{pmatrix}
\bar{\theta}_n \\ \theta^n 
\end{pmatrix} \,.
\end{align}

The light-cone form of the $C$-matrix before a gauge fixing is 
\begin{align}
C = 
\begin{pmatrix}
t_a{}^n \; | \; s_{an}  
\end{pmatrix}
\,.
\end{align}
A priori, the $GL(k)_L$ index $a$ is not correlated with the light-cone index $n$. We choose to fix the gauge by locking $GL(k)_L$ and $GL(k)_R$:
\begin{align}
C = 
\begin{pmatrix}
t_a{}^n \; | \; s_{an}  
\end{pmatrix}
\quad 
\rightarrow 
\quad 
\begin{pmatrix}
\delta_m{}^n \; | \; u_{mn}  
\end{pmatrix} \,, 
\quad 
u_{mn} = (t^{-1})_m{}^a s_{an} \,.
\end{align}
The orthogonality condition implies that $u_{mn}$ is anti-symmetric:
\begin{align}
\delta(C\cdot g\cdot C^T ) = \delta\left(\thalf(u_{mn}+u_{nm})\right) \,.
\end{align}
Since the decomposition of $u_{mn}$ into the symmetric and anti-symmetric parts is a linear operation, 
the delta-function does not produce any $u$-dependent Jacobian factor. 

We will call this gauge fixing the ``u-gauge". This gauge was inspired by the fact that the orthogonal Grassmannian and the pure spinor admit the same 
$SO(2k)/U(k)$ coset description and that the $u_{mn}$ coordinates were used in ref.~\cite{Berkovits:2004bw} to solve the non-linear constraints of the pure spinors in order to construct higher dimensional twistor transforms. 

It is well known that $C\cdot \lambda =0$ and $C\cdot g \cdot C^T=0$ implies the overall momentum conservation. In the light-cone gauge, $C\cdot \lambda =0$ is 
written as 
\begin{align}
w_m + u_{mn} v^n = 0 \,.
\label{w-uv}
\end{align}
This equation admits a particular form of $SL(k,\mathbb{R})$-invariant solution for all $k$:
\begin{align}
u_{mn}^* = \frac{2\langle w_m w_n \rangle}{\langle w_p v^p \rangle} 
= \frac{1}{R} \langle w_m w_n \rangle \,, 
\quad 
R \equiv \frac{\langle w_p v^p \rangle}{2} \,.
\label{u-star}
\end{align}
To verify that \eqref{u-star} is indeed a solution to \eqref{w-uv}, 
it suffices to use the Schouten identity
\begin{align}
\langle w_m w_n \rangle v^{p\alpha} + 
\langle w_n v^p \rangle w_m^\alpha +
\langle v^p w_m \rangle w_n^\alpha = 0 \,,
\end{align}
and the momentum conservation \eqref{p-cons}. 

The light-cone basis before the gauge fixing 
respects the symmetry exchanging $w_n$ and $v^n$. 
Thus it is natural to consider the ``dual u-gauge" 
in which the roles of $w_n$ and $v^n$ are reversed:
\begin{align}
C = 
\begin{pmatrix}
x_a{}^n \; | \; u_{an}  
\end{pmatrix}
\quad 
\rightarrow 
\quad 
\begin{pmatrix}
\bar{u}^{mn} \; | \; \delta^m{}_n   
\end{pmatrix} \,.
\end{align}
In the dual u-gauge, the $C\cdot \lambda = 0$ condition reads
\begin{align}
v^m + \bar{u}^{mn} w_n=0 \,,
\end{align}
which admits a particular solution, 
\begin{align}
\bar{u}^{mn}_* = - \frac{1}{R}  \langle v^m v^n \rangle \,.
\end{align}
The dual u-gauge will be useful in a later discussion on the $\lambda$-parity for odd $k$.
Using the energy momentum conservation and Schouten identity, one can show that
\begin{align}
u_{mp}^* \bar{u}^{np}_* = -\frac{1}{R} \langle w_m v^n \rangle \,,
\quad 
u_{mp}^* \bar{u}^{mp}_* = -2 \,.
\label{uubar}
\end{align}

For $k=2$ and $k=3$, \eqref{u-star} is the unique solution to \eqref{w-uv}. 
For higher $k$, there is a $(k-2)(k-3)/2$-dimensional solution space 
containing \eqref{u-star}. For instance, for $k=4$, 
the general solution can be parametrized by
\begin{align}
\hat{u}_{mn}(z) = u_{mn}^* + 
\frac{z}{2} \epsilon_{mnrs} \bar{u}_*^{rs} \,.
\end{align}
The general solution for $k=5$ is 
\begin{align}
\hat{u}_{mn}(z) = u_{mn}^* + 
\frac{z^p}{2} \epsilon_{mnprs} \bar{u}_*^{rs} \,. 
\end{align}
The ``vector" $z^p$ appears to have five components, but only three of them are independent due to the equivalence relation, 
\begin{align}
z^p \sim z^p + c_\alpha v^{p\alpha}\,,
\end{align}
which follows from the fact that $\bar{u}_*^{rs} \propto \langle v^m v^n \rangle$ and the Schouten identity. 
Along the same line of reasoning, 
we can write the general solution for $k\ge 4$ as
\begin{align}
\hat{u}_{mn}(z) = u_{mn}^* +  \frac{1}{2(k-4)!} z^{p_1 \cdots p_{k-4}}\epsilon_{mn p_1 \cdots p_{k-4} rs} \bar{u}_*^{rs} \,.
\label{u-star-general}
\end{align} 
In the $(k-2)$-dimensional space surviving the quotient $z^p \sim z^p + c_\alpha v^{p\alpha}$, the tensor $z^{p_1 \cdots p_{k-4}}$ spans a $(k-4)$-plane. 
The effective number of components for $z^{p_1 \cdots p_{k-4}}$ is, as expected,  
\begin{align}
\binom{k-2}{k-4} = \binom{k-2}{2}  =\frac{(k-2)(k-3)}{2}\,.
\end{align} 

\subsubsection*{Broader class of u-gauges}

Most features of the u-gauge survives permutations of columns. For instance, 
instead of the alternating signature metric \eqref{split-sign}, we may take
\begin{align}
g = \mathrm{diag}(\underbrace{-,\cdots,-}_k,\underbrace{+,\cdots,+}_k)\,,
\label{minus-plus}
\end{align}
and define light-cone coordinates by 
\begin{align}
w_m = (-1)^{k-1} (x_{m+k} + x_{m})\,, \quad v^m = (-1)^{k-1} (x_{m+k} - x_{m}) \,.
\label{minus-plus-2}
\end{align}
One of the useful features of all $u$-type gauges is that the Jacobian factors arising from the computation of amplitudes are always powers of the $R$ factor 
defined in \eqref{u-star}, although the numerical value of $R$ does depend on the particular gauge. 

There are $(2k)!/(k!)^2$ different ways to distribute $(-1)$ and $(+1)$ in the diagonal entries of the metric. An overall flip of the signs is irrelevant, 
so there are $(2k-1)!/(k!(k-1)!)$ inequivalent metrics. 
Given a fixed metric, there are $k!$ inequivalent ways to pair the coordinates to define light-cone coordinates. To sum up, the number of different u-type gauges is $(2k-1)!/(k-1)!$. 

Among all possibilities, we will mostly focus on the two choices we mentioned explicitly above. Both of them generalizes to arbitrary $k$ straightforwardly.  
The alternating signature gauge defined \eqref{split-sign}, \eqref{split-sign-2} is the only choice which respects the cyclic symmetry. For this reason we will call this gauge ``u-cyclic gauge". As we will see later, the other gauge defined by \eqref{minus-plus}, \eqref{minus-plus-2} is convenient when we examine the factorization of $\mathcal{A}_{2k}$ into two copies of $\mathcal{A}_{k+1}$ when $k$ is odd. We will call this choice ``u-factorization gauge". 

\subsubsection*{Lambda-parity in the u-gauge}

Let us examine how the lambda parity is reflected in the u-gauge. 
We will show that, for odd $k$, the lambda parity induces 
the exchange, 
\begin{align}
w_m \;\; \leftrightarrow \;\; v^m \,.
\label{uv-exchange}
\end{align}
For notational convenience, we will work in the u-factorization gauge, 
but the same arguments hold in all u-type gauges. 

With a usual gauge fixing in the particle basis \cite{Gang:2010gy,Lee:2010du}, the $C$-matrix is given by
\begin{align}
C = 
\begin{pmatrix}
I \; | \; O 
\end{pmatrix}  \,,
\qquad 
O O^T = I \,.
\end{align}
In the light-cone basis, the $C$-matrix translates to
\begin{align}
\hat{C} = 
\begin{pmatrix}
I - O \; | \; I + O  
\end{pmatrix}  \,.
\end{align}
Note that the following identities hold for odd-dimensional orthogonal matrices:
\begin{align}
\det(O) = \pm 1 
\quad 
\Longrightarrow
\quad
\det(I \mp O) = 0 \,.
\end{align}
When $\det(O)=-1$, a $GL(k)$ gauge transformation gives rise to 
\begin{align}
\hat{C} \rightarrow
\begin{pmatrix}
I \; | \; (I-O)^{-1}(I + O)  
\end{pmatrix} 
= 
\begin{pmatrix}
I \; | \; U 
\end{pmatrix} \,.
\end{align}
This establishes the relation between the u-gauge and the usual gauge 
in the particle basis. 

The other branch with $\det(O)=+1$ is related to the $\det(O)=-1$ branch 
by $\det(-O) = -\det(O)$. The sign flip has the same effect 
as flipping the signs of all $\lambda_m^\alpha$ for $m=k+1,\cdots,2k$. 
Up to an overall $SO(2k)$ rotation, this is the same as the exchange 
\eqref{uv-exchange}. 
Thus we have proved that the lambda parity induces 
the exchange of $w_m$ and $v^m$.

\section{Tree amplitudes}

\subsection{4-point}

The momentum conservation in the particle basis reads, 
\begin{align}
P = 
-\lambda_1 \lambda_1 +\lambda_2 \lambda_2 -\lambda_3 \lambda_3
+\lambda_4 \lambda_4 = 0\,,
\end{align}
where we suppressed the spinor indices. In terms of the Lorentz scalars, 
$\langle ij\rangle$, we obtain 
\begin{align}
&\frac{\langle 12 \rangle}{\langle 34 \rangle}
= \frac{\langle 23 \rangle}{\langle 14 \rangle}
= - \frac{\langle 31 \rangle}{\langle 24 \rangle} 
= \sigma = \pm 1\,, 
\label{4pt-branch}
\\
& \langle 12 \rangle^2 + \langle 23 \rangle^2 - \langle 31 \rangle^2 = 0 \,.
\label{4pt-cp1}
\end{align}
The sign factor $\sigma$ in \eqref{4pt-branch} specifies a branch of OG$_4$. 
Without loss of generality, we will work in the $\sigma=+1$ branch 
for the rest of this subsection. 

In the light-cone basis, 
the gauge-fixed $C$-matrix and the metric are $(u= u_{12})$
\begin{align}
\widehat{C} = 
\begin{pmatrix}
1 & 0 & 0 & u \\
0 & 1 & -u & 0  
\end{pmatrix}\,, 
\quad
\hat{g} = \frac{1}{2}
\begin{pmatrix}
0 & 0 & 1 & 0 \\
0 & 0 & 0 & 1 \\
1 & 0 & 0 & 0 \\
0 & 1 & 0 & 0 \\
\end{pmatrix} \,.
\end{align}
In the particle basis, 
\begin{align}
C = 
\begin{pmatrix}
1 & 1 & -u & u \\
u & -u & 1 & 1 
\end{pmatrix}\,, 
\quad
g = \mathrm{diag}(-,+,-,+)\,.
\end{align}
To avoid confusion, 
we put hats on the objects in the light-cone basis.

In the evaluation of the Grassmannian integral, 
the kinematic delta-function gives 
\begin{align}
\delta(C\cdot \lambda) = J_4^B \, \delta^3(P) \,\delta(u-u_*) \,. 
\label{4pt-kin}
\end{align}
The value of $u_*$ is determined by \eqref{u-star}:
\begin{align}
u_* 
= \frac{2\langle w_1 w_2 \rangle}{\langle w_1 v^1 \rangle+\langle w_2 v^2 \rangle}
= \frac{\langle 23 \rangle - \langle 31 \rangle }{\langle 12 \rangle} 
= -\frac{\langle 12 \rangle}{\langle 23 \rangle  +\langle 31 \rangle} \,.
\end{align}
The equality of three expressions follow from \eqref{4pt-branch} 
with $\sigma=+1$ and \eqref{4pt-cp1}. 
The Jacobian factor in \eqref{4pt-kin} is 
\begin{align}
J_4^B = \frac{ \langle v^1 v^2 \rangle } {2} = - (\langle 23 \rangle + \langle 31 \rangle ) \,.
\end{align}
The fermionic delta function gives
\begin{align}
\delta(C\cdot \eta) = J_4^F \delta^6(Q) = \left( \frac{4}{\langle v^1 v^2\rangle} \right)^3 \delta^6(Q) \,.
\end{align}
The denominator at $u=u_*$ is 
\begin{align}
\left. \frac{1}{M_1M_2} \right|_{u=u_*} = \frac{1}{(-2u_*)(1-u_*^2)} 
= \frac{(\langle 23 \rangle + \langle 31 \rangle )^2}{4 \langle 12 \rangle \langle 23 \rangle} \,.
\end{align}
Collecting all ingredients, we reproduce the standard form of the 4-point amplitude, 
\begin{align}
\mathcal{A}_4 = \frac{\delta^3(P)\delta^6(Q)}{\langle 12 \rangle \langle 23 \rangle} \,.
\label{4pt-final}
\end{align}

\subsection{6-point}

It is well known that the Grassmannian integral for the 6-point amplitude is fully localized by the delta functions and leaves no contour integral. In the particle basis, the gauge-fixed $C$ matrix in the $(+)$-branch is
\begin{align}
C^+ = 
\begin{pmatrix}
1 & 1 & -u_{12} & u_{12} & -u_{13} & u_{13} \\
u_{12} & -u_{12} & 1 & 1 & -u_{23} & u_{23} \\
u_{13} & -u_{13} & u_{23} & -u_{23} & 1 & 1
\end{pmatrix}\,, 
\quad
g = \mathrm{diag}(-,+,-,+,-,+)\,.
\end{align}
The kinematic delta-function can be transformed into
\begin{align}
\delta(C\cdot \lambda) = J_6^B \, \delta^3(P) \,\delta^3(u_{mn}-u_{mn}^*) \,. 
\label{6pt-kin}
\end{align}
The value of $u_{mn}^*$ is determined by \eqref{u-star} and the Jacobian factor is $J_6^B=1/2$.
 The fermionic delta function gives
\begin{align}
\delta(C\cdot \eta) = J_6^F \delta^6(Q) \delta^3(\zeta_+)  \,,
\quad 
\zeta_{+}^I = \frac{1}{2} \epsilon^{mnp} u_{mn}^* \bar{\theta}_p^I \,.
\end{align}
The Jacobian factor 
from the fermionic delta function is 
\begin{align}
J_6^F = \left(\frac{4}{R}\right)^3\,,
\end{align}
with $R$ defined in \eqref{u-star}.
The denominator at $u_{ij}=u_{ij}^*$ is 
\begin{align}
\left. \frac{1}{M_1^+M_2^+M_3^+} \right|_{u_{mn}=u_{mn}^*} = \frac{1}{8(u_{13}^*-u_{12}^*u_{23}^*)(u_{23}^*-u_{12}^*u_{13}^*)(u_{12}^*-u_{13}^*u_{23}^*)} \,.
\end{align}
Collecting all ingredients, we obtain the result 
of the Grassmannian integral in the u-gauge:
\begin{align}
\frac{4\, \delta^3(P) \delta^6(Q) \delta^3(\zeta_{+})}{R^3 (u_{13}^*-u_{12}^*u_{23}^*)(u_{23}^*-u_{12}^*u_{13}^*)(u_{12}^*-u_{13}^*u_{23}^*)} \,.
\label{6pt-half}
\end{align}
The full amplitude is the sum of the contributions from the two branches 
related to each other by $\lambda$-parity. 
As explained in the previous section, for odd $k$, 
$\lambda$-parity exchanges the two types of light-cone coordinates. 
In terms of the super-space variables, the exchange means 
\begin{align}
w_m \leftrightarrow v^m \,,
\quad \bar{\theta}_{m}\leftrightarrow \theta^{mI} \,, 
\quad 
\zeta_+^I \leftrightarrow \zeta_-^I = \frac{1}{2}\epsilon_{mnp} \bar{u}^{mn}_* \theta^{pI} \,.
\end{align}
The explicit form of the $C$-matrix in the conjugate branch is
\begin{align}
C^- &= 
\begin{pmatrix}
-1 & 1 & \bar{u}^{12} & \bar{u}^{12} & \bar{u}^{13} & \bar{u}^{13} \\
-\bar{u}^{12} & -\bar{u}^{12} & -1 & 1 & \bar{u}^{23} & \bar{u}^{23} \\
-\bar{u}^{13} & -\bar{u}^{13} & -\bar{u}^{23} & -\bar{u}^{23} & -1 & 1
\end{pmatrix}\,.
\end{align}
Summing up the two terms, we obtain the full 6-point amplitude:
\begin{align}
\mathcal{A}_{6}=&\delta^3(P) \delta^6(Q) 
\frac{4}{R^3} 
\left(
\frac{\delta^3(\zeta_{+})}{(u_{13}^*-u_{12}^*u_{23}^*)(u_{23}^*-u_{12}^*u_{13}^*)(u_{12}^*-u_{13}^*u_{23}^*)} \right.
\nn \\
&\qquad\qquad\qquad\qquad 
\left. + \frac{\delta^3(\zeta_{-})}{(\bar{u}^{13}_*-\bar{u}^{12}_*\bar{u}^{23}_*)(\bar{u}^{23}_*-\bar{u}^{12}_*\bar{u}^{13}_*)(\bar{u}^{12}_*-\bar{u}^{13}_*\bar{u}^{23}_*)}
\right)  \,,
\label{6pt-result-a}
\end{align}

The planar 6-point amplitude can be factorized by two 4-point amplitudes in three different channels. At first sight, it is not clear 
how the result \eqref{6pt-result-a} can exhibit the factorization properties. 
Remarkably, it is possible to show that the consecutive minors from the two branches combine to produce the desired physical poles. In the u-gauge, 
we have 
\begin{align}
M_1^+ M_1^- = \left( \frac{4}{R} \right)^2 p_{123}^2 \,, 
\quad
M_2^+ M_2^- = - \left( \frac{4}{R} \right)^2 p_{234}^2 \,, 
\quad
M_3^+ M_3^- = \left( \frac{4}{R} \right)^2 p_{345}^2 \,,  
\end{align}
where $p_{ijk \cdots}^2 = (p_i +p_j +p_k + \cdots )^2$. 
A proof of this relation and discussion 
on its gauge (in)dependence is presented in appendix \ref{sec:MM}.

\subsection{8-point \label{sec:8pt}}

\subsubsection*{Contour integral}
As discussed in section \ref{sec:u-gauge}, the general solution 
to $C\cdot\lambda=0$ in a u-gauge is 
\begin{align}
\hat{u}_{mn}(z) = u^*_{mn}+\frac{z}{2}\epsilon_{mnpq}\bar{u}^{pq}_* \,.
\end{align}
The Grassmannian integral reduces to a contour integral in $z$ 
through the relation
\begin{align}
\delta^{8}(C\cdot\lambda)=J^B_8\; \delta^3(P) \int dz \; \delta^6(u_{mn}-\hat{u}_{mn}(z))\,, 
\end{align}
with $J^B_8 = 1/(2R)$. Up to an overall sign, the full 8-point amplitude is obtained
when the contour separates the poles of $M_1$ and $M_3$ 
from those of $M_2$ and $M_4$ \cite{Gang:2010gy}. 

The minors of $C$-matrix can be at most quartic in $\hat{u}_{mn}(z)$. But, explicit computations show that all quartic terms can be absorbed into the square of the quadratic polynomial, 
\begin{align}
\hat{u}_{mn}(z)\hat{u}_{pq}(z) + \hat{u}_{mp}(z)\hat{u}_{qn}(z) + \hat{u}_{mq}(z)\hat{u}_{np}(z) = - z \,\epsilon_{mnpq} \,. 
\label{u2-comb}
\end{align}
Similarly, all cubic terms can be rewritten as the same polynomial \eqref{u2-comb} times a linear combination of $\hat{u}_{mn}(z)$. 
These two statements imply that all minors of $C$, including the consecutive ones, are quadratic in $z$:
\begin{align}
M_i(z) = a_i z^2+ b_i z + c_i = a_i(z-z_i^+)(z-z_i^-)\,.
\end{align}
The fermionic delta function produces 
\begin{align}
\delta^{12}(C\cdot \eta) = \left( \frac{4}{R}\right)^3 \delta^6(Q) \prod_{I=1}^3 (A_I z^2 + B_I z + C_I) \,,
\end{align}
where the fermion bilinears $(A_I,B_I,C_I)$ are defined as follows: 
\begin{align}
&\qquad \textstyle{\frac{1}{4}} \epsilon^{mnpq} u^*_{mn} 
\left(\bar{\theta}^I_p +\thalf z \epsilon_{prxy}\bar{u}^{xy}_*\theta^{rI} \right) \left(\bar{\theta}^I_q +\thalf z\epsilon_{qszw}\bar{u}^{zw}_*\theta^{sI} \right) 
\nonumber \\
&=\textstyle{\frac{1}{4}} \left[ -z^2\epsilon_{mnpq}\bar{u}_*^{mn}\theta^{pI}\theta^{qI} + z\epsilon^{mnpq} \epsilon_{prxy}\bar{u}^*_{mn} \bar{u}_*^{xy} \theta^{rI} \bar{\theta}_q^I  + \epsilon^{mnpq}u^*_{mn}\bar{\theta}_p^I\bar{\theta}_q^I \right]
\nonumber \\ 
&= A_I z^2 +B_I z+C_I \,.
\end{align}

We wish to evaluate the contour integral 
\begin{align}
I(\mathcal{C}) = \oint_\mathcal{C} \frac{dz}{2\pi i} \frac{\prod_{I=1}^3 (A_I z^2 + B_I z + C_I)}{\prod_{i=1}^4(a_i z^2+ b_i z + c_i) } \,.
\label{8pt-int}
\end{align}
Note that, as far as the evaluation of the integral is concerned, 
the distinction between $(a_i,b_i,c_i)$ and $(A_I, B_I,C_I)$ 
is immaterial. To streamline notations, we replace 
$(A_I,B_I,C_I)$ $(I=1,2,3)$ by $(a_i,b_i,c_i)$ $(i=5,6,7)$ in what follows. 
To illustrate the ideas behind the computation, we find it useful to 
introduce a family of integrals of similar form, 
\begin{align}
I_n(\mathcal{C}) = \oint_\mathcal{C} \frac{dz}{2\pi i} \frac{\prod_{j=n+2}^{2n+1} (a_j z^2 + b_j z + c_j)}{\prod_{i=1}^{n+1}(a_i z^2+ b_i z + c_i) } \,.
\label{n-int}
\end{align}
We can recover the original integral \eqref{8pt-int} by choosing $n=3$.

These integrals share two crucial features. One is that they are homogeneous functions of the variables $(a_i,b_i,c_i)$ with degree $(-1)$ for 
$i=1, \ldots, n+1$ and $(+1)$ for $i=n+2,\ldots,2n+1$.  
The other is that they are invariant under the $SL(2,\mathbb{C})$ transformation, 
\begin{align}
z \rightarrow \frac{e z +f}{gz+h} \,,
\quad 
\begin{pmatrix}
e & f \\ g & h 
\end{pmatrix} 
\in SL(2,\mathbb{C}) \,.
\end{align}
It is instructive to consider the generators of $SL(2,\mathbb{C})$ one by one:
\begin{align} 
\mbox{Dilatation:}&\quad z \rightarrow e^{t} z \,,
\\
\mbox{Inversion:}&\quad z \rightarrow -1/z \,,
\\
\mbox{Translation:}&\quad z \rightarrow z + \epsilon \,.
\end{align}
The change in $z$ can be reproduced exactly by the change in the coefficients:
\begin{align}
\mbox{Dilatation:}&\quad (a_i,b_i,c_i) \rightarrow (e^{t} a_i, b_i, e^{-t}c_i ) \,,
\\
\mbox{Inversion:}&\quad (a_i,b_i,c_i) \rightarrow (c_i, -b_i, a_i ) \,,
\\
\mbox{Translation:}&\quad (a_i,b_i,c_i) \rightarrow (a_i, b_i + 2a_i \epsilon, c_i + b_i \epsilon + a_i \epsilon^2 ) \,. 
\end{align}
%
The integral \eqref{n-int} should be invariant under the $SL(2,\mathbb{C})$ action on $z$, provided that the contour transforms accordingly. 
It follows that the result of the integral should be invariant 
under the change of coefficients listed above. 
	
To be specific, let us focus on the contribution of the contour $\mathcal{C}_1$ enclosing the two poles $z_1^{\pm}$ only. The residue theorem gives 
\begin{align}
I_n(\mathcal{C}_1) &= \frac{1}{a_1} \left( \frac{N_n(z_1^+)}{D_n(z_1^+)} - \frac{N_n(z_1^-)}{D_n(z_1^-)} \right) \frac{1}{z_1^+ - z_1^-} 
\nn \\
&= \frac{1}{a_1 D_n(z_1^+)D_n(z_1^-)} \cdot \frac{N_n(z_1^+) D_n(z_1^-) - N_n(z_1^-)D_n(z_1^+)}{z_1^+ - z_1^-} 
\,,
\end{align}
where we defined 
\begin{align}
N_n(z) = \prod_{j=n+2}^{2n+1} M_j(z) \,,
\quad 
D_n(z) = \prod_{i=2}^{n+1} M_i(z)\,, 
\quad M_i(z) = a_i z^2+ b_i z + c_i\,. 
\end{align}
The product $D_n(z_1^+)D_n(z_1^-)$ is easy to evaluate. Using the relations
\begin{align}
z_1^+ + z_1^- = - \frac{b_1}{a_1} \,,
\quad 
z_1^+ z_1^- = \frac{c_1}{a_1} \,,
\end{align}
we find 
\begin{align}
D_n(z_1^+)D_n(z_1^-) = a_1^{-2n} \prod_{i=2}^{n+1} \Delta_{1i} \,,
\end{align}
where we defined short-hand notations 
\begin{align}
&\Delta_{ij} = \beta_{ij}^2 - \alpha_{ij} \gamma_{ij}\,,
\quad
\alpha_{ij} = b_i c_j - b_j c_i\,, 
\;\;
\beta_{ij} = c_i a_j - c_j a _i  \,,
\;\; 
\gamma_{ij} = a_i b_j - a_j b_i \,.
\end{align}
The new symbols $(\alpha_{ij}, \beta_{ij}, \gamma_{ij})$ 
obey simple $SL(2,\mathbb{C})$ transformation rules, 
\begin{align}
\mbox{Dilatation:}&\quad (\alpha_{ij},\beta_{ij},\gamma_{ij}) \rightarrow (e^{-t} \alpha_{ij}, \beta_{ij}, e^{t} \gamma_{ij} ) \,,
\\
\mbox{Inversion:}&\quad (\alpha_{ij},\beta_{ij},\gamma_{ij}) \rightarrow (\gamma_{ij}, -\beta_{ij}, \alpha_{ij} ) \,,
\\
\mbox{Translation:}&\quad (\alpha_{ij},\beta_{ij},\gamma_{ij}) \rightarrow (\alpha_{ij} -2\beta_{ij}\epsilon  + \gamma_{ij} \epsilon^2, \beta_{ij} -\gamma_{ij} \epsilon, \gamma_{ij}) \,, 
\end{align}
so that $\Delta_{ij}$ is fully invariant under $SL(2,\mathbb{C})$. 

To summarize what we have done so far, 
\begin{align}
I_n(\mathcal{C}_1) 
= \frac{1}{\prod_{i=2}^{n+1} \Delta_{1i}} \cdot (a_1)^{2n-1} \frac{N_n(z_1^+) D_n(z_1^-) - N_n(z_1^-)D_n(z_1^+)}{z_1^+ - z_1^-} 
\,,
\end{align}
The remaining $z_1^{\pm}$-dependent part may look complicated 
as both $N_n(z)$ and $D_n(z)$ are degree $2n$ polynomials in $z$. 
However, since we only need their values at the two solutions of $M_1(z)=0$, 
we can take the polynomial quotients.  
If we denote the quotient and the remainder by
\begin{align}
N_n(z) = Q_n(z) M_1(z) + R_n z + S_n \,,
\quad
D_n(z) = \tilde{Q}_n(z) M_1(z) + \tilde{R}_n z + \tilde{S}_n \,,
\label{poly-quo}
\end{align}
the integral gives
\begin{align}
I_n(\mathcal{C}_1) 
= \frac{(a_1)^{2n-1} (R_n \tilde{S}_n - S_n \tilde{R}_n) }{\prod_{i=2}^{n+1} \Delta_{1i}} \,.
\end{align}
The denominator $\prod_{i=2}^{n+1} \Delta_{1i}$ has degree $(2n;2,\cdots,2;0,\cdots,0)$. 
It remains to express the numerator $F_n\equiv (a_1)^{2n-1} (R_n \tilde{S}_n - S_n \tilde{R}_n)$, which has degree $(2n-1;1\cdots,1;1,\cdots,1)$, in an $SL(2,\mathbb{C})$ invariant way. 
To do so, we introduce a few additional $SL(2,\mathbb{C})$-invariants:
\begin{align}
J_{ijk} &= \det(a_i,b_j,c_k) = 6a_{[i}b_jc_{k]} \,, 
\nn \\
K_{ij} &= b_ib_j - 2 (a_i c_j +  c_i a_j) \,,
\nn \\
L_{ijkl} &= \alpha_{ij} \gamma_{kl} - 2 \beta_{ij} \beta_{kl} + \gamma_{ij} \alpha_{kl} \,.
\end{align}

For $n=0$, the integral vanishes trivially as the contour can be pushed to infinity without encountering any poles. For $n=1$, the numerator 
$F_n$ should be of degree $(1;1;1)$ in three groups of variables 
and anti-symmetric with respect to the last two. 
It appears that $J_{123}$ is the only $SL(2,\mathbb{C})$-invariant 
with required properties. An explicit computation indeed shows that 
\begin{align}
F_1 = -J_{123}\,.
\end{align}
For $n=2$, we look for a polynomial of degree $(3;1,1;1,1)$ 
with total symmetry under permutations in the same group and 
anti-symmetry between the last two groups. 
The answer indeed respects all the desired properties: 
\begin{align}
F_2 = \frac{1}{2} K_{1(2} J_{3)1(4} K_{5)1} 
+ \frac{1}{2} K_{11} J_{1(2}{}^{(4} K_{3)}{}^{5)} \,.
\label{F2}
\end{align}
Finally, we turn to $n=3$, our original problem. 
There are a number of ways to combine $J_{ijk}$, $K_{ij}$, $L_{ijkl}$ to construct $SL(2,\mathbb{C})$-invariants 
with correct symmetry properties. Remarkably, the answer 
can be organized using only two such combinations: 
\begin{align}
F_3 = - K_{11} J_{12}{}^{(5} J_{13}{}^{6} J_{14}{}^{7)}
+\frac{3}{4} L^1{}_{(2}{}^1{}_3 J_{4)1(5} L^1{}_{6}{}^1{}_{7)} \,.
\label{F3}
\end{align}

A remark is in order. 
The integrals $I_n(\mathcal{C})$ are defined in such a way that 
if we set, say, $M_3(z)=M_5(z)$, $I_2(\mathcal{C}_1)$ should reduce to $I_1(\mathcal{C}_1)$. 
In terms of $F_n$, we should have 
\begin{align}
F_2(12345)|_{``3=5"} = \Delta_{13} F_1(124)\,.
\end{align}
The reduction does not look obvious from the expression \eqref{F2}. 
Similarly, it is not obvious how the reduction from $I_3$ to $I_2$ occurs:
\begin{align}
F_3(1234567)|_{``4=7"} = \Delta_{14} F_2(12356)\,.
\end{align}
It is conceivable that the decompositions \eqref{F2} and \eqref{F3} 
are not unique, and some alternative decomposition will 
make the reduction more obvious. 

\subsubsection*{8-point amplitude: the result}

In summary, the 8-point amplitude can be written as 
\begin{align}
\mathcal{A}_{8}= \delta^3(P) \delta^6(Q)   (1+\pi) J_8^B J_8^F  \left(\frac{F(1)}{\Delta_{12} \Delta_{13} \Delta_{14}}+\frac{F(3)}{\Delta_{31} \Delta_{32} \Delta_{34}} \right)  \,,
\label{8pt-final}
\end{align}
with the Jacobian factors
\begin{align}
J_8^B  = \frac{1}{2R} \,,
\quad 
J_8^F = \left(\frac{4}{R}\right)^3 \,.
\end{align}
With the $\lambda$-parity operator $\pi$, the $(1+\pi)$ factor denotes the sum over two branches of the orthogonal Grassmannian.
In \eqref{8pt-final}, we removed the subscript from $F_3$ and and added reference to the contour by $F(i)$. Note that this form of the 8-point amplitude is valid in any u-type gauges.

\subsubsection*{Physical and spurious poles}

The 8-point amplitude \eqref{8pt-final} is the sum of two contour integrals, 
$I(\mathcal{C}_1)$ encircling the poles from $M_1$ and $I(\mathcal{C}_3)$ encircling the poles from $M_3$. Each term 
carries physical and spurious poles. The most convenient tool to analyze the pole structure is the on-shell diagram pioneered by \cite{ArkaniHamed:2012nw} and elaborated for ABJM amplitudes in \cite{Huang:2013owa,Kim:2014hva,Huang:2014xza}.

\begin{figure}[htbp]
	\centering
	\includegraphics[width=9cm]{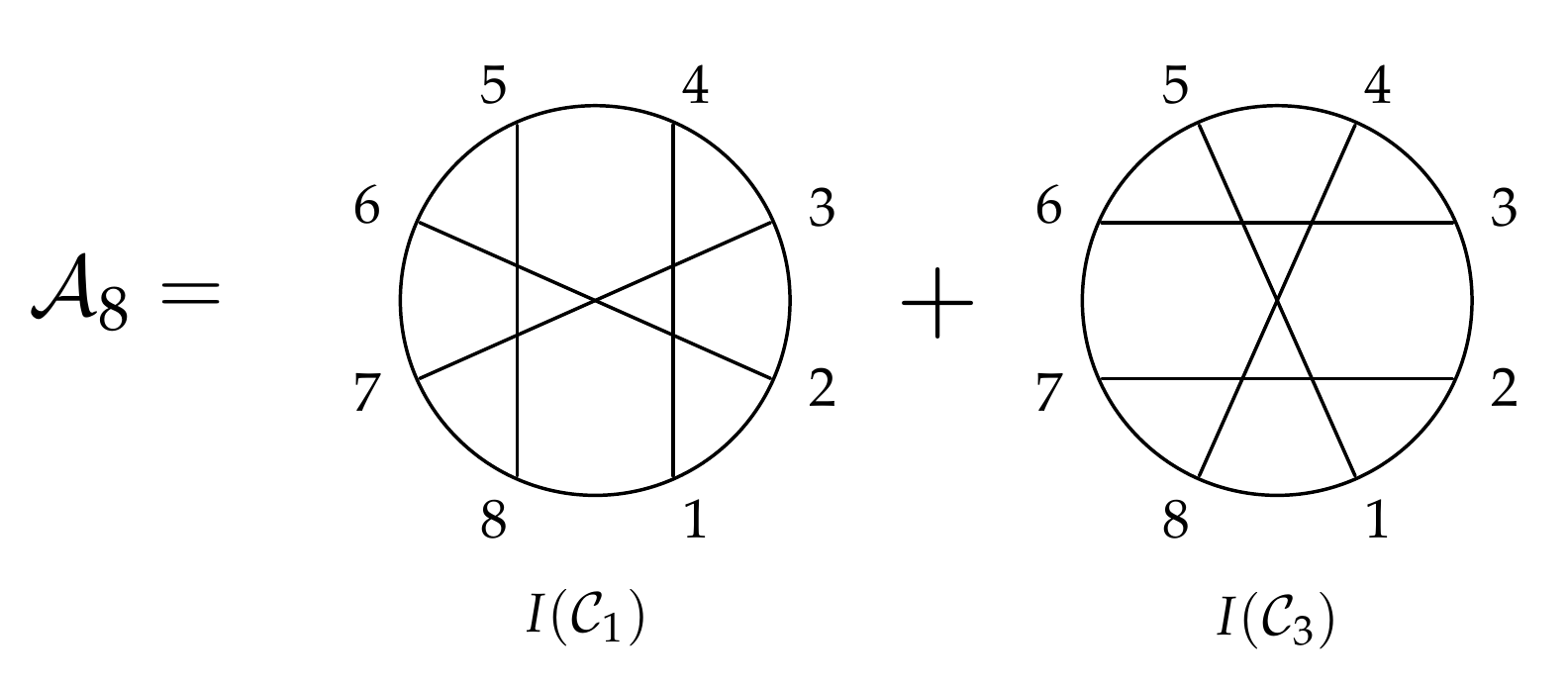}
\caption{On-shell digrams for the 8-point amplitude collect residues from $M_1$ and $M_3$.}
	\label{8pt-bcfw}
\end{figure}

The on-shell diagram of ABJM amplitudes are built from two building blocks: a quartic vertex and an internal line. The quartic vertex is precisely the 4-point amplitude \eqref{4pt-final}, and the internal line equates the kinetic variables of two legs from different vertices and integrates ($\int d^{2|3}\Lambda$).

\begin{figure}[htbp]
	\centering
	\includegraphics[width=14.4cm]{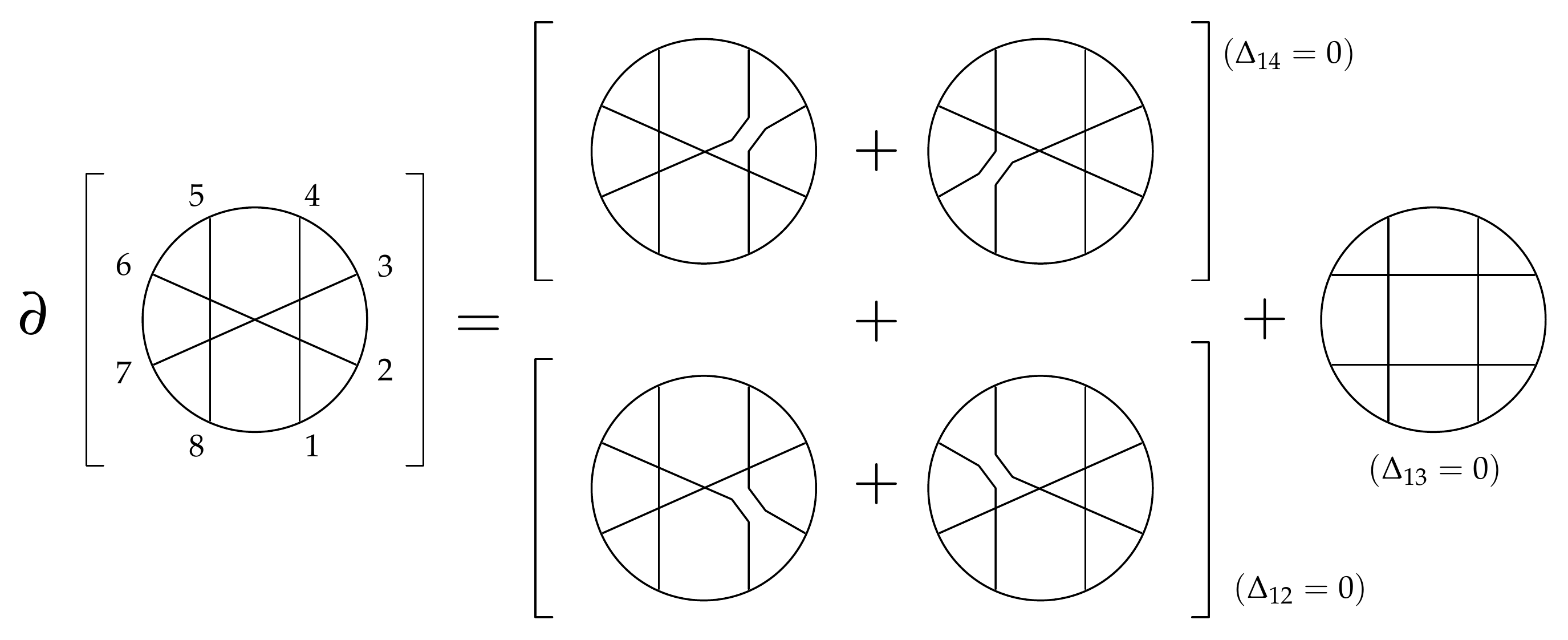}
\caption{Poles of $I(C_1)$ correspond to boundary components of the on-shell diagram.}
	\label{8pt-poles}
\end{figure}

The poles of the amplitude corresponds to 
boundaries of the on-shell diagrams. 
Each on-shell diagram has five vertices. 
Barring disconnected diagrams, each vertex yields exactly one boundary term.  
Figure~\ref{8pt-poles} shows the five boundary terms from the on-shell diagram 
for $I(\mathcal{C}_1)$. 
Using the canonical coordinates for on-shell diagrams introduced in \cite{Huang:2013owa,Kim:2014hva,Huang:2014xza}, we can easily see which consecutive minors vanish as we approach each of the five boundary components. To be specific, we adopt the coordinates of 
\cite{Kim:2014hva} associated with the OG tableaux.

\begin{figure}[htbp]
	\centering
	\includegraphics[width=8.5cm]{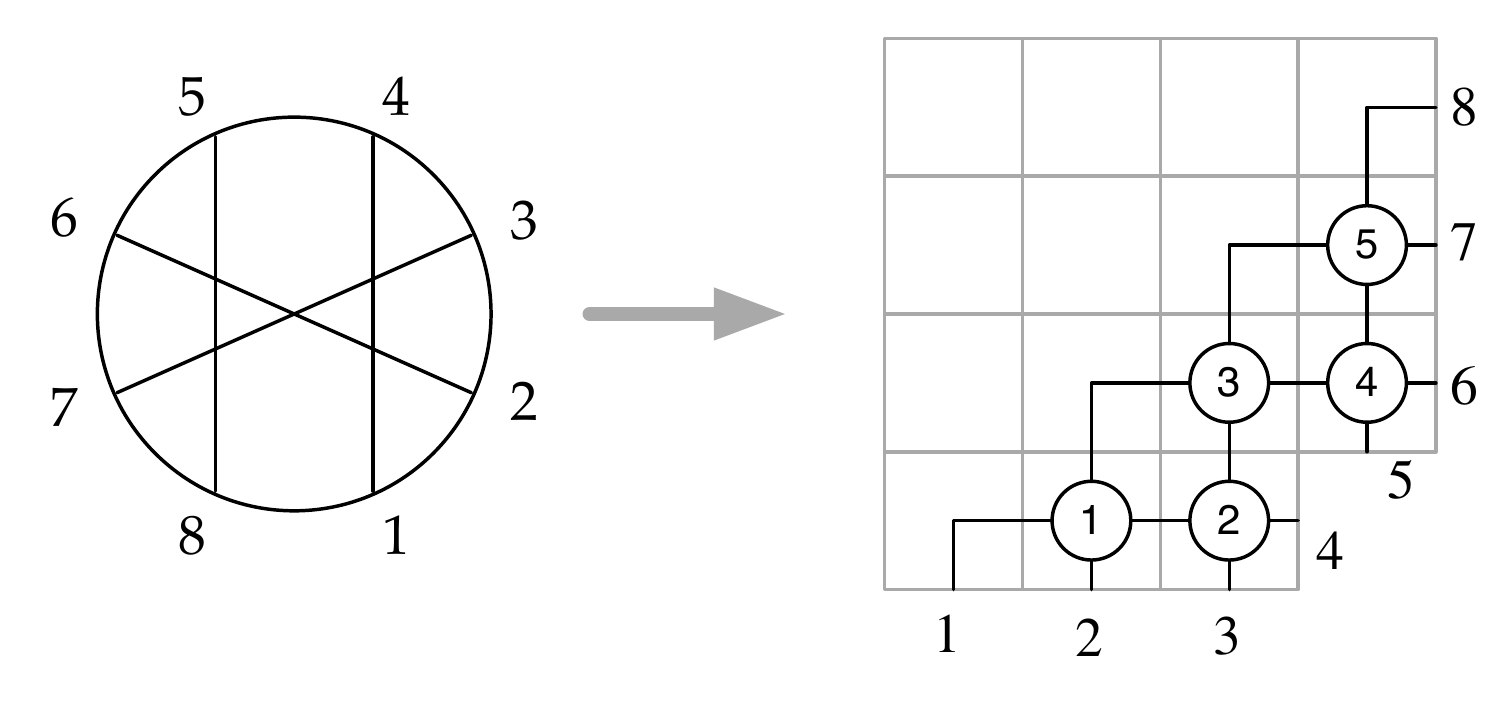}
\caption{OG tableaux for $I(\mathcal{C}_1)$.}
	\label{8pt-tableaux}
\end{figure}

\begin{figure}[htbp]
	\centering
	\includegraphics[width=10cm]{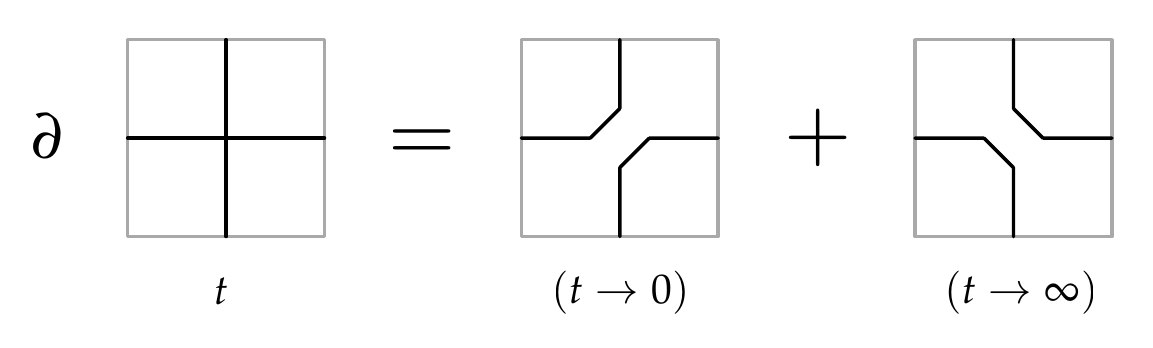}
\caption{Boundary operation in terms of canonical coordinates.}
	\label{8pt-boundary}
\end{figure}

The tableau for $I(\mathcal{C}_1)$ is depicted in Figure~\ref{8pt-tableaux}. 
It can be translated to the $C$-matrix according to the rules explained in \cite{Kim:2014hva}. Let $C_i$ be the $i$-th column of the $C$-matrix. We begin by setting the `source' columns $(C_1,C_2,C_3,C_5)$ to form an identity matrix. 
We assign a coordinate $t_v$ to each vertex. To fill in the `sink' columns $(C_4,C_6,C_7,C_8)$, 
we consider all paths from a source to a sink which may move upward and to the right but not downward or to the left. The path picks up $\pm \sinh(t_v)$ if 
it passes through the vertex, or $\pm \cosh(t_v)$ if it makes a turn at the vertex. The final matrix element is given by a polynomial of the form, schematically, 
\begin{align}
C_{\mathrm{source},\mathrm{sink}} = \sum_{\mathrm{paths}} \prod_\mathrm{vertices} f_v \,.
\end{align}
We refer the readers to \cite{Kim:2014hva} for details. 
All we need here is the remarkable fact that the consecutive minors are 
given by {\em monomials} of the $\sinh(t)$ factors.:
\begin{align}
M_1 = 0 \,,
\quad 
M_2 = s_1 s_2 \,,
\quad 
M_3 = s_2 s_3 s_4 \,,
\quad 
M_4 = s_4 s_5 \,, 
\quad 
s_v \equiv \sinh(t_v)\,.
\end{align}
In these coordinates, the boundary operation amounts to taking one of the 
coordinate variables to zero or infinity. The orientation 
of the untied diagram in the OG tableaux is shown in Figure~\ref{8pt-boundary}. 

To compare with the factorization channels in Figure~\ref{8pt-poles}, 
it is convenient to rescale the minors by an overall factor, $\widetilde{M}_i = M_i/(s_2 s_4)$, 
\begin{align}
\widetilde{M}_1 = 0 \,,
\quad 
\widetilde{M}_2 = \frac{s_1}{s_4} \,,
\quad 
\widetilde{M}_3 = s_3 \,,
\quad 
\widetilde{M}_4 = \frac{s_5}{s_2} \,.
\end{align}
The rescaled minor $\widetilde{M}_4$ vanishes in the limit $s_2\rightarrow \infty$ or $s_5\rightarrow 0$. Through the prescriptions in Figure~\ref{8pt-tableaux} and \ref{8pt-boundary}, the two limits give the two boundary diagrams on top of Figure~\ref{8pt-poles}, which in turn corresponds to the factorization 
channels for $p_{123}^2$ and $p_{567}^2$, respectively. 

In the contour integral obtained earlier, the simultaneous vanishing of 
$M_1$ and $M_4$, or equivalently the `collision' of poles from $M_1$ and $M_4$, 
would result in the vanishing of $\Delta_{14}$. It is then natural to expect that $\Delta_{14}$, a polynomial of kinematic variables, 
is proportional to $p_{123}^2 p_{567}^2$. 
In the u-gauge, we can explicitly verify the proportionality between $\Delta_{ij}$ 
and physical poles. By symmetry, we expect that all of the eight physical poles are indeed associated with ``collision" of roots of the minors:
\begin{align}
p_{123}^2 p_{567}^2 \propto R^4\Delta_{14} \,, \;\;
p_{234}^2 p_{678}^2 \propto R^4\Delta_{12}\,, \;\; 
p_{345}^2 p_{781}^2 \propto R^4\Delta_{32} \,, \;\; 
p_{456}^2 p_{812}^2 \propto R^4\Delta_{34} \,. 
\label{phys-poles}
\end{align}
The powers of $R$ are fixed on dimensional ground. 
We leave the details of the verification, including the numerical coefficients,  to appendix \ref{sec:NN}.

We can identify the poles for $\Delta_{13} = \Delta_{31}$ in \eqref{8pt-final} as spurious poles. A standard argument in the Grassmannian integral 
uses the fact that 
\[
\mathcal{A}_8 = I(\mathcal{C}_1) + I(\mathcal{C}_3) = - I(\mathcal{C}_2) - I(\mathcal{C}_4) \,.
\] 
Since $\Delta_{13} = \Delta_{31}$ arises from $I(\mathcal{C}_1)$ and 
$I(\mathcal{C}_3)$ but not from $I(\mathcal{C}_2)$ or $I(\mathcal{C}_4)$, 
it must be spurious. The physical poles \eqref{phys-poles}, in contrast, appear in both contour prescriptions. A related observation is that the on-shell diagram 
for $\Delta_{13}=0$ in Figure~\ref{8pt-poles} can cancel against 
the same diagram from the boundary of $I(\mathcal{C}_3)$ if sign factors are properly assigned. 

We conclude this section with a few remarks on the generalization of the methods we used. The u-gauge has some advantages over more familiar gauges based on Euler angles. One of them is the decomposition of the fermionic delta-function, 
\begin{align}
&\delta^{3k}(C\cdot \eta) = J_{2k}^F \delta^{6}(Q) \prod_I \zeta^I\,, 
\quad J_{2k}^F = \left(\frac{4}{R}\right)^3 \,,
\nn \\
&\zeta^I = \frac{1}{2(k-2)!}\epsilon^{mnp_1\cdots p_{k-2}} u^*_{mn} 
(\bar{\theta}_{p_1}^I+\hat{u}_{p_1q_1}\theta^{q_1 I})\cdots 
(\bar{\theta}_{p_{(k-2)}}^I+\hat{u}_{p_{(k-2)}q_{(k-2)}}\theta^{q_{(k-2)}I}) \,,
\end{align} 
with $\hat{u}_{pq}$ linear in the $z$ coordinates in \eqref{u-star-general}. 
Another advantage is that, as explained in appendix \ref{sec:NN}, 
the minors take a relatively simple form in the u-gauge. 

Finally, in anticipation of the generalization to 10-point or higher amplitudes, we note that the $SL(2,\mathbb{C})$ invariants are related to cross-ratios. For instance, 
\begin{align}
\frac{\Delta_{14}}{K_{11} K_{44}} = 
\frac{(z_{1+} - z_{4+})(z_{1-} - z_{4-})(z_{1+} - z_{4-})(z_{1-} - z_{4+})}{(z_{1+}-z_{1-})^2 (z_{4+}-z_{4-})^2} \,.
\end{align}
Higher point amplitudes would inevitably give rise to more complicated ``collision of poles" and it would be crucial to introduce higher dimensional analogs of $\Delta, J, K, L$ invariants to work without explicitly solving quadratic equations for the $z$ coordinates.


\section{Soft theorem for ABJM amplitudes}

Soft theorems in gauge (gravity) theories explore the limit in which one or more gluon (graviton) approaches vanishing momenta. It is well-known that the soft limit of a non-vanishing tree amplitude is divergent and that the leading divergent term takes a universal form. More recently the sub-leading terms in the soft limit were calculated by using on-shell techniques \cite{Cachazo:2014fwa} and spurred renewed interest in soft theorems and their applications. 
In this section, we derive a soft theorem for ABJM tree amplitudes, following a similar analysis for three-dimensional supergravity theories \cite{Chen:2014cuc} (see also \cite{Huang:2013owa} for an early consideration of the double soft limit of ABJM theory).

Since the ABJM amplitudes are well-defined only for even number of external particles, it is natural to define the double soft limit of the $(2k+2)$-point amplitude $\mathcal{A}_{2k+2}$ by scaling the momenta of the last two particles, 
\begin{align}
(p_{2k+1},p_{2k+2})  &\rightarrow \epsilon^2 (p_{2k+1}, p_{2k+2}) \,,
\end{align}
and taking the $\epsilon \rightarrow 0$ limit. In spinor variables, 
the scaling rule is 
\begin{align}
(\lambda_{2k+1}, \lambda_{2k+2}) 
&\rightarrow
\epsilon (\lambda_{2k+1} , \lambda_{2k+2}) \,.
\end{align}
In view of the soft theorems in gauge theories in various dimensions,
we anticipate that $\mathcal{A}_{2k+2}$ in the soft limit reduces to the $\mathcal{A}_{2k}$ up to a universal soft factor $S(\epsilon)$,
\begin{align}
\mathcal{A}_{2k+2}|_{\epsilon \rightarrow 0} = S(\epsilon) \mathcal{A}_{2k} \,.
\end{align}
We will find that the soft factor consists of a leading and a sub-leading term: 
\begin{align}
S(\epsilon) = \frac{1}{\epsilon^2} S^{(0)}+ \frac{1}{\epsilon} S^{(1)} \,.
\end{align} 

\subsection{Recursion relation for soft limit}

Following the approach of ref.~\cite{Chen:2014cuc}, we will use the BCFW recursion relation for ABJM amplitudes  to analyze the double soft theorem. It is convenient to choose the two reference particle in the BCFW recursion to be $(2k)$ and $(2k+1)$, namely, neighboring soft and hard particles. The BCFW-shifted kinematic variables are given by 
\begin{align}
\label{soft-bcfw}
&\hat{\lambda}_{2k} = c \lambda_{2k} + s \lambda_{2k+1} \,, \quad\;\;\; \hat{\eta}_{2k} = c \eta_{2k} + s \eta_{2k+1} \,,    \nn \\
&\hat{\lambda}_{2k+1} = s \lambda_{2k} + c \lambda_{2k+1}\,, \quad  \hat{\eta}_{2k+1} = s \eta_{2k} + c \eta_{2k+1} \,.
\end{align}
where $c=\cosh t$ and $s=\sinh t$ with $c^2-s^2 = 1$ and $z \equiv c+s = e^t$.

\begin{figure}
\centering
\begin{tikzpicture}
\node at (0,0) [circle,draw,fill=black!20!white,minimum size=1cm] {L};
\node at (3,0) [circle,draw,fill=black!20!white,minimum size=2cm] {R};
\draw (0.5,0) -- (2,0); \draw (1.25, 0.3) node {$p_{f}$} ;
\draw (0,0.5) -- (0,1); \draw (0,1.4) node {$\widehat{2k+1}$} ;
\draw (-0.5,0) -- (-1,0) ;  \draw (-1.7,0) node {$2k+2$} ;
\draw (0,-0.5) -- (0,-1);  \draw (0 , -1.4) node {$1$} ;
\draw (3,1) -- (3,1.6); \draw (3,1.9) node {$\widehat{2k}$} ;
\draw (3,-1) -- (3,-1.6); \draw (3,-1.9) node {$2$} ;
\draw[dashed](3,1.4) arc (90:-90:1.4cm);
\end{tikzpicture}
\caption{Factorization diagram}
\label{fig:factdig}
\end{figure}
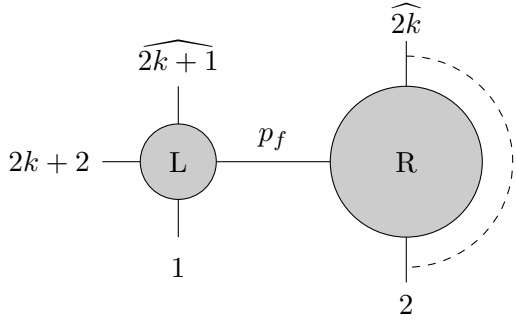

As explained in \cite{Cachazo:2014fwa,Chen:2014cuc} for soft graviton theorems, 
only one of the terms in the BCFW recursion formula contributes to the divergent soft factors. In our notation, the term is depicted 
in Figure~\ref{fig:factdig}. Let us briefly review why this is the case. 
The recursion formula schematically takes the form:
\begin{align}
\mathcal{A}_{2k+2} = \sum_{l=2}^{k} \hat{\mathcal{A}}^{(L)}_{2l} \frac{H(z_+,z_-)}{p_f^2} \hat{\mathcal{A}}^{(R)}_{2(k+2-l)}\,,
\end{align}
where $\hat{\mathcal{A}}^{(L)}_{2l}$ and $\hat{\mathcal{A}}^{(R)}_{2(k+2-l)}$ are two sub-amplitudes evaluated at BCFW-deformed momenta, $p_f$ is the momentum through the factorization channel, and 
\begin{align}
H(x,y) = \frac{x^{2}(y^{2}-1)}{x^{2}-y^{2}} 
\end{align}
is the BCFW kernel introduced in ref.~\cite{Gang:2010gy}.  

When $l=2$ as in Figure~\ref{fig:factdig}, in the soft limit $(p_{2k+1},p_{2k+2})  \rightarrow \epsilon^2 (p_{2k+1}, p_{2k+2})$, $p_f$ in Figure~\ref{fig:factdig} becomes nearly equal to $p_1$ such that $1/p_f^2 \approx 1/p_1^2$ diverges by the on-shell condition of $p_1$. For $l>2$ with generic assignment of external momenta, $1/p_f^2$ remains finite. The other factors $\hat{\mathcal{A}}^{(L)}_{2l}$, $\hat{\mathcal{A}}^{(R)}_{2(k+2-l)}$, $H(z_+,z_-)$ also remain finite. 

Focusing on the term in Figure~\ref{fig:factdig}, 
we need to evaluate the following amplitude
\begin{align}
\mathcal{A}_{2k+2}|_{\epsilon \rightarrow 0}  
&= \int d^{3} \eta_{f} \left[ \mathcal{A}_{4}(\hat{\Lambda}_{2k+1}, \Lambda_{2k+2}, \Lambda_{1}, \hat{\Lambda}_{f})(z_{+})
\frac{H(z_{+}, z_{-})}{p_{1,2k+1,2k+2}^{2}} \mathcal{A}_{2k}( \hat{\Lambda}_{f}, \cdots, \hat{\Lambda}_{2k})(z_{+}) \right]
\nonumber \\
&\qquad + (z_+ \leftrightarrow z_-) 
\nn \\
&\equiv \mathcal{A}_{2k+2}^{(+)} + \mathcal{A}_{2k+2}^{(-)}\,.
\label{bcfw-soft}
\end{align}

In this diagram, the BCFW-deformed $p_f$ is given by 
\begin{align}
\hat{p}_{f}(z) &= p_{2} + p_{3}+ \cdots + p_{2k-1} +\hat{p}_{2k}(z) 
= - p_{1} -p_{2k+2} -\hat{p}_{2k+1}(z) \,.
\end{align}	
The on-shell values $z_+$, $z_-$ are the two solutions of 
the on-shell condition  $\hat{p}_{f}(z)^{2}=0$.
We can solve this condition order by order in $\epsilon$ by assuming that $s=s_{0}+\epsilon s_{1}+\epsilon^{2} s_{2} +\cdots $ and $c=c_{0}+\epsilon c_{1}+\epsilon^{2} c_{2} +\cdots$, with $c^{2}-s^{2}=1$. The solutions are given by \cite{Chen:2014cuc} 
\begin{align}
c_\pm &=1-\frac{\epsilon^{2}}{2}\alpha_{\pm}^2+\mathcal{O}(\epsilon^{4}) \,, \\
s_\pm &= - \epsilon \alpha_{\pm}+\frac{\epsilon^{3}}{4}\left[ (\alpha_{\pm}+\alpha_{\mp}) \alpha_{\pm}^{2} - (\alpha_{\pm}-\alpha_{\mp})\beta_{\pm}^{2} \right] +\mathcal{O}(\epsilon^{5}) \,,\\
z_\pm & = 1-\epsilon \alpha_\pm - \frac{\epsilon^2}{2}\alpha_\pm^2 +\mathcal{O}(\epsilon^3)
\end{align}
where $\alpha_{j}$ and $\beta_{j}$ are defined by
\begin{align}
\alpha_{\pm} = \frac{\langle 1,2k+1 \rangle \pm \langle 1,2k+2 \rangle}{\langle 1,2k \rangle} \,, 
\quad 
\beta_{\pm} =\frac{\langle 2k,2k+1 \rangle \pm \langle 2k,2k+2 \rangle}{\langle 1,2k \rangle} \,.
\end{align}
As mentioned in \cite{Chen:2014cuc}, the $(+)$ solution corresponds to $\sigma = +1$, i.e.
\begin{align}
\langle 1,f\rangle =\langle \widehat{2k+1}, 2k+2\rangle, \quad 
\langle f, \widehat{2k+1}\rangle= \langle 1, 2k+2\rangle, \quad 
\langle f,2k+2\rangle = - \langle \widehat{2k+1}, 1\rangle, 
\end{align}
and the $(-)$ solution corresponds to $\sigma=-1$, i.e.,
\begin{align}
\langle 1,f\rangle =-\langle \widehat{2k+1}, 2k+2\rangle, \;\;
\langle f, \widehat{2k+1}\rangle= -\langle 1, 2k+2\rangle, \;\; 
\langle f,2k+2\rangle = \langle \widehat{2k+1}, 1\rangle.
\end{align}
These relations will be useful when we calculate the soft limit. 
 
Note that, in \eqref{soft-bcfw}, the first correction terms carry $\epsilon^2$ weight relative to the leading terms for bosonic variables, whereas the relative weight is $\epsilon^{1}$ for fermionic variables. In order to compute the leading and sub-leading terms of the soft limit, we need only the leading correction terms for bosonic variables, but we should keep track of leading and next to leading corrections for fermions.  

We first focus on the $\mathcal{A}^{(+)}_{2k+2}$ contribution to \eqref{bcfw-soft}. 
Upon expansion in powers of $\epsilon$, the propagator in the factorization channel gives
\begin{align}
\frac{1}{(p_{1}+p_{2k+1}+p_{2k+2})^{2}} =-\frac{1}{\epsilon^{2}} \frac{1}{\alpha_{+}\alpha_{-}\langle 1,2k \rangle^{2}} \,,
\end{align}
and the BCFW kernel gives 
\begin{align}
H(z_{+},z_{-}) = \frac{z_{+}^{2}(z_{-}^{2}-1)}{z_{+}^{2}-z_{-}^{2}} = \frac{\alpha_{-}}{\alpha_{+}-\alpha_{-}}  \,.
\end{align}
The bosonic factors in the 4-point amplitude reduce to 
\begin{align}
\langle f , \widehat{2k+1} \rangle &= \epsilon \langle 1 , 2k+2 \rangle \,, 
\nn \\
\langle f , 2k+2 \rangle &= - \langle \widehat{2k+1} , 1 \rangle =  -\epsilon \langle 1,2k+2 \rangle \,, 
\nn \\
\langle f , 1 \rangle &= \langle 2k+2 , \widehat{2k+1} \rangle = \epsilon^{2} \langle 1 , 2k+2 \rangle \beta_{+}\,.
\end{align}
The fermionic delta function in the 4-point amplitude can be rewritten as
\begin{align}
\delta^{6}(Q) &= \delta^{3}\left(\eta_{f}+\frac{\langle \widehat{2k+1}, 2k+2 \rangle }{\langle \widehat{2k+1} ,f \rangle} \eta_{2k+2}
- \frac{\langle \widehat{2k+1}, 1 \rangle }{\langle \widehat{2k+1}, f \rangle} \eta_{1}\right) 
\nonumber \\
& \quad \times \delta^{3}(-\langle f ,\widehat{2k+1} \rangle \hat{\eta}_{2k+1} +\langle f ,2k+2 \rangle \eta_{2k+2}-\langle f , 1 \rangle \eta_{1}) 
\nn \\
&= \delta^3(\eta_f + \chi) \times (-\epsilon^3 \langle 1 , 2k+2 \rangle^3) \times \delta^3(\eta_{2k+1} + \eta_{2k+2}+ \epsilon(-\alpha_+ \eta_{2k} +\beta_+ \eta_1)) \,.
\end{align}
The first factor absorbs the fermionic integral in \eqref{bcfw-soft}. The remaining factors combine with the bosonic factors to produce $\mathcal{O}(1)$ and $\mathcal{O}(\epsilon^1)$ terms in the soft limit. 

\begin{figure}[htbp]
	\centering
	\includegraphics[width=11cm]{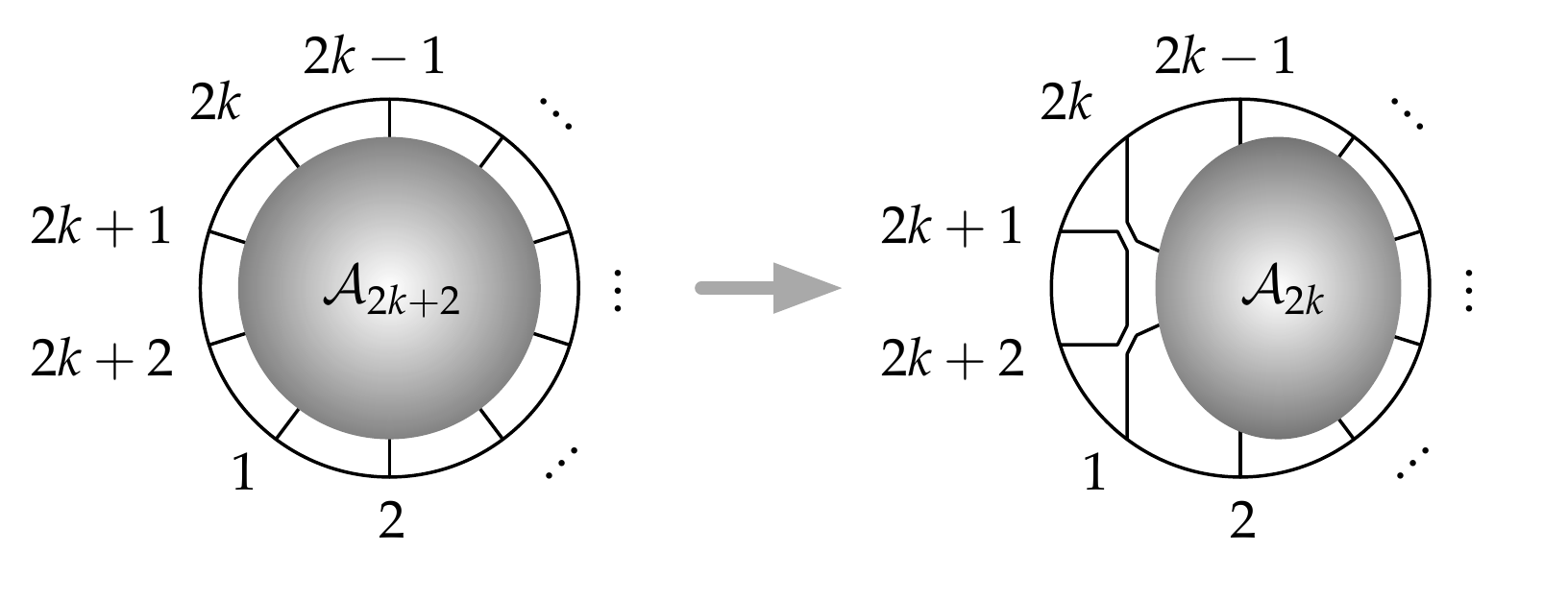}
\caption{Soft limit from the on-shell diagram perspective.}
	\label{soft-onshell}
\end{figure}

Collecting all ingredients, we find the $z_+$ contribution to the 
soft limit of $\mathcal{A}_{2k+2}$: 
\begin{align}
\mathcal{A}_{2k+2}^{(+)} 
= \frac{\delta^3(\tilde{\eta}_{2k+1} + \tilde{\eta}_{2k+2})}{2\epsilon^2 \langle 1,2k\rangle \alpha_+ \beta_+}  
\mathcal{A}_{2k}(\tilde{\eta}_1, \eta_2 ,\eta_3, \cdots, \eta_{2k-1}, \tilde{\eta}_{2k}) \,,
\end{align} 
where
\begin{align}
\begin{pmatrix}
\tilde{\eta}_{1} \\ \tilde{\eta}_{2k+2}
\end{pmatrix}
= 
\begin{pmatrix}
1 & \epsilon \beta_+ \\
\epsilon \beta_+ & 1 
\end{pmatrix}
\begin{pmatrix}
\eta_{1} \\ \eta_{2k+2}
\end{pmatrix}
\,,
\quad 
\begin{pmatrix}
\tilde{\eta}_{2k} \\ \tilde{\eta}_{2k+1}
\end{pmatrix}
= 
\begin{pmatrix}
1 & -\epsilon \alpha_+ \\
-\epsilon \alpha_+ & 1 
\end{pmatrix}
\begin{pmatrix}
\eta_{2k} \\ \eta_{2k+1}
\end{pmatrix} \,.
\end{align}
Expanding explicitly in powers of $\epsilon$, we obtain 
%
\begin{align}
\mathcal{A}_{2k+2}^{(+)}
&=\frac{\delta^3(\bar{\theta}_{k+1})}{2\epsilon^2 \langle 1,2k\rangle \alpha_+\beta_+}  \mathcal{A}_{2k}
\nonumber \\
&\quad +\frac{1}{2\epsilon \langle 1,2k\rangle \alpha_+\beta_+}\left[  
\thalf \epsilon_{IJK}\bar{\theta}_{k+1}^I\bar{\theta}_{k+1}^J \xi_+^K 
+  \delta^3(\bar{\theta}_{k+1})\left(\beta_+ R_{2k+2,1} -\alpha_+ R_{2k+1,2k}  \right)  \right] \mathcal{A}_{2k} \,.
\label{A_+}
\end{align}
where we introduced 
\begin{align}
\xi_+ = -\alpha_+\eta_{2k} +\beta_{+}\eta_{1} \,,
\quad
R_{i,j} = \eta_{i} \frac{\partial}{\partial \eta_j} \,.
\end{align}
Note that $R_{i,j}$ are $R$-symmetry generators of the ABJM theory. 
Note also that we could have obtained exactly the same result if we had chosen external particles $(2k+2)$ and $(1)$ as the reference legs for the BCFW recursion. In this sense, the symmetry between $(2k,2k+1)$ and $(2k+2,1)$ has been restored. This is natural from the on-shell diagram perspective as illustrated in Figure~\ref{soft-onshell}.

The computation of $\mathcal{A}^{(-)}_{2k+2}$ in \eqref{bcfw-soft} proceeds in the same way. The result is 
\begin{align}
\mathcal{A}_{2k+2}^{(-)}
&=-\frac{\delta^3(\theta_{k+1})}{2\epsilon^2 \langle 1,2k\rangle \alpha_-\beta_-}  \mathcal{A}_{2k}
\nonumber \\
&\quad +\frac{1}{2\epsilon \langle 1,2k\rangle \alpha_- \beta_-}\left[  
\thalf \epsilon_{IJK} \theta_{k+1}^I \theta_{k+1}^J \xi_-^K 
+  \delta^3(\theta_{k+1})\left(\beta_- R_{2k+2,1} + \alpha_- R_{2k+1,2k}  \right)  \right] \mathcal{A}_{2k} \,.
\label{A_-}
\end{align}
with
$\xi_- = -\alpha_-\eta_{2k} +\beta_{-}\eta_{1}$. 
Note that $\mathcal{A}_{2k+2}^{(+)}$ and  $\mathcal{A}_{2k+2}^{(-)}$ are related to each other by lambda-parity, $\Lambda_{2k+2}\rightarrow-\Lambda_{2k+2}$, 
in addition to the exchange $(\alpha_+,\beta_+) \leftrightarrow (\alpha_-,\beta_-)$.

Finally, we add the two contributions to obtain the leading and the sub-leading soft factor of the double soft limit
\begin{align}
\mathcal{A}_{2k+2}|_{\epsilon \rightarrow 0}
= \left(\frac{1}{\epsilon^2} S^{(0)} + \frac{1}{\epsilon} S^{(1)} \right) \mathcal{A}_{2k} + \mathcal{O}(1) \,.
\end{align}
where the leading and sub-leading soft factors are
\begin{align}
S^{(0)} &= \frac{1}{2\langle 1,2k \rangle} \left[
	\frac{\delta^3(\bar{\theta}_{k+1})}{ \alpha_+\beta_+} 
	-\frac{\delta^3(\theta_{k+1})}{\alpha_-\beta_-} 
	 \right] \,,\\
S^{(1)} &= \frac{1}{2 \langle 1,2k\rangle \alpha_+\beta_+}\left[  
\thalf \epsilon_{IJK}\bar{\theta}_{k+1}^I\bar{\theta}_{k+1}^J \xi_+^K 
+  \delta^3(\bar{\theta}_{k+1})\left(\beta_+ R_{2k+2,1} -\alpha_+ R_{2k+1,2k}  \right)  \right]
\nonumber \\
&\quad +\frac{1}{2 \langle 1,2k\rangle \alpha_- \beta_-}\left[  
\thalf \epsilon_{IJK} \theta_{k+1}^I \theta_{k+1}^J \xi_-^K 
+  \delta^3(\theta_{k+1})\left(\beta_- R_{2k+2,1} + \alpha_- R_{2k+1,2k}  \right)  \right] \,.
\end{align}

\subsection{Soft limit of the 6-point amplitude}
For simplicity, we will use $\lambda$-parity operator $\pi$ which is given by
\begin{align}
\pi \; : \; \Lambda_6 \rightarrow - \Lambda_6 \,.
\end{align}
If we use this operator, we can consider only one part of 6-point amplitude. 

The soft limit with particle 5 and 6 soft is realized in the light-cone basis as
\begin{align}
w_{3} \rightarrow \epsilon \, w_{3} \,, \quad 
v^{3} \rightarrow \epsilon \, v^{3} \,.
\end{align}
As we observed earlier, the bosonic kinematic invariants receive leading corrections at the $\epsilon^2$ order. So we can freely use the 4-point kinematic relations. For example, 
\begin{align}
R &= \frac{1}{2}(\langle w_1 v^1 \rangle + \langle w_2 v^2 \rangle +\langle w_3 v^3 \rangle)   
= \langle 12 \rangle +  \langle 34 \rangle + \mathcal{O}(\epsilon^2) = 2\langle 12 \rangle +  \mathcal{O}(\epsilon^2)\,, \\
-1&= u_{12}^* \bar{u}^{12}_*  + u_{13}^* \bar{u}^{13}_* +  u_{23}^* \bar{u}^{23}_* 
=u_{12}^* \bar{u}^{12}_* + \mathcal{O}(\epsilon^2)  \,.
\end{align}
In the soft limit, up to $\mathcal{O}(\epsilon^2)$ terms, the minors become 
\begin{align}
M_1^+ &= 2 (u_{13}^* - u_{12}^* u_{23}^*) 
= 2u_{12}^* (- \bar{u}^{12}_* u_{13}^*-u_{23}^* ) 
= \frac{2u_{12}^*}{R} (\langle w_3 v^2 \rangle + \langle w_3 w_2 \rangle ) 
\nn \\
&=  \frac{2u^*_{12}}{R} (- 2\epsilon) ( \langle 45 \rangle + \langle 46 \rangle )  
= \epsilon \left( - \frac{4 \langle 14 \rangle u^*_{12}}{R} \right) \beta_+ \,,  
\\
M_2^+ &= -2 (u_{23}^* - u_{12}^* u_{13}^*) = 2u_{12}^* (\bar{u}^{12}_* u_{23}^*+ u_{13}^* ) = \frac{2u_{12}^*}{R} (- \langle v^1 w_3 \rangle + \langle w_1 w_3 \rangle ) 
\nn \\
& =  \frac{2u^*_{12}}{R} (2\epsilon)  (\langle 15 \rangle + \langle 16 \rangle) = \epsilon \left( \frac{4 \langle 14 \rangle u^*_{12}}{R} \right) \alpha_+ \,,
\\
M_3^+ & = - 2(u_{12}^* -u_{13}^*u_{23}^*) = - 2 u_{12}^* \,.
\end{align}
Recall that the 6-point amplitude \eqref{6pt-result-a} contains two fermionic parts \begin{align}
 \delta^3(\zeta_+) \,, \quad \delta^6(Q_6) \,.
\end{align}
Neglecting $\mathcal{O}(\epsilon^2)$ terms, we observe that 
\begin{align}
\delta^3(\zeta_+) 
= (u_{12}^*)^3 \delta^{3}\left( \bar{\theta}_3 + \frac{u_{23}^*}{u_{12}^*}\bar{\theta}_1 -\frac{u_{13}^*}{u_{12}^*} \bar{\theta}_2 \right) 
= (u_{12}^*)^3 \delta^{3}( \bar{\theta}_3 + \epsilon(-\alpha_+ \eta_4 + \beta_+ \eta_1)) \,.
\label{softzeta}
\end{align}
The second identity follows from 
\begin{align}
\frac{u_{23}^*}{u_{12}^*}\bar{\theta}_1 -\frac{u_{13}^*}{u_{12}^*} \bar{\theta}_2  
&=  -{u_{23}^*}{\bar{u}^{12}_*} \bar{\theta}_1 + u_{13}^*\bar{u}^{12}_* \bar{\theta}_2 
= - \frac{1}{R} \left( \langle w_3 v^1 \rangle  \bar{\theta}_1 + \langle w_3 v^2 \rangle \bar{\theta}_2 \right)
\nn \\
&= 
- \frac{ 1}{2 \langle 12 \rangle} \left[ \langle w_3 | ( | 2 \rangle - | 1 \rangle ) ( \eta_2+\eta_1 )+ \langle w_3 | ( | 4 \rangle - | 3 \rangle ) ( \eta_4+\eta_3 ) \right]
\nn \\
&= 
- \frac{ 1}{2 \langle 12 \rangle}  \langle w_3 | ( | 2 \rangle \eta_1 - | 1 \rangle \eta_2 + | 4 \rangle \eta_3  - | 3 \rangle  \eta_4 ) 
\nn \\
&=  \epsilon(-\alpha_+ \eta_4 + \beta_+ \eta_1) \,,
\label{softzeta2}
\end{align}
where we used (super)-momentum conservation 
\begin{align}
- |1 \rangle \eta_1 +|2 \rangle \eta_2- |3 \rangle \eta_3 + |4 \rangle \eta_4 
= \mathcal{O}(\epsilon) \,, 
\quad 
\langle 14 \rangle = \langle 23 \rangle  + \mathcal{O}(\epsilon^2) \,,
\nn 
\end{align}
and Schouten identities. 

We now move on to the $\delta^6(Q_6)$ factor. To check our result, it is better to start with our conjecture. From our recursion relation result, $\mathcal{A}_{4}$ part gives the super-momentum conservation like 
\begin{align}
\widetilde{Q}_{4}
&\equiv - | 1 \rangle (\eta_1 + \epsilon  \beta_+ \eta_6 ) + | 2 \rangle \eta_2 - | 3 \rangle \eta_3 + | 4 \rangle ( \eta_4 - \epsilon \alpha_+ \eta_5) 
\nn \\
&= - |1\rangle \left(\eta_{1} +\epsilon \frac{\langle 45 \rangle + \langle 46 \rangle}{\langle 14 \rangle}\eta_{6} \right)
+ |2 \rangle \eta_{2} - |3 \rangle \eta_{3}
+ |4\rangle  \left(\eta_{4} - \epsilon \frac{\langle 15 \rangle + \langle 16 \rangle}{\langle 14 \rangle}\eta_{5} 
\right) 
\nonumber \\
&= Q_{4} - \frac{\epsilon}{\langle 14 \rangle} 
\left(
\langle 15 \rangle |4\rangle \eta_{5} + \langle 46 \rangle |1\rangle \eta_{6}
+\langle 16 \rangle |4\rangle \eta_{5} + \langle 45 \rangle |1\rangle \eta_{6} 
\right) 
\nonumber \\
&= Q_{4} - \frac{\epsilon}{\langle 14 \rangle} \left(
 \left( \langle 14 \rangle | 5 \rangle + \langle 45 \rangle | 1 \rangle  \right) \eta_{5}
- \left(\langle 14 \rangle | 6 \rangle - \langle 16 \rangle |4 \rangle \right) \eta_{6} 
+ \langle 16 \rangle |4\rangle \eta_{5} 
+ \langle 45 \rangle |1\rangle \eta_{6}
\right) 
\nonumber \\
&= Q_{6}
- \epsilon\frac{\langle 45 \rangle}{\langle 14 \rangle} |1\rangle (\eta_{5} + \eta_{6})
- \epsilon\frac{\langle 16 \rangle}{\langle 14 \rangle} |4\rangle (\eta_{5} + \eta_{6}) \nonumber\\
&= Q_{6}	\,.
\end{align}
The last equality holds on the support of \eqref{softzeta}. 
So we can conclude that the six-point supermomentum conservation becomes the four-point supermomentum conservation with next-leading soft correction.
Finally, our 6-point amplitude becomes
\begin{align}
\mathcal{A}_6|_{\epsilon \rightarrow 0} 
& = (1+\pi) \left( \frac{32 \delta^3(P) \delta^6(Q_6) \delta(\zeta_+)}{R^3 M_1^+ M_2^+ M_3^+} \right)  \nn \\
& = (1+\pi) \left( \frac{\delta^3(\bar{\theta}_3 + \epsilon(-\alpha_+ \eta_4 + \beta_+ \eta_1))}
{2 \epsilon^2 \langle 14 \rangle \alpha_+\beta_+} \frac{ \delta^3(P) \delta^6 (\hat{Q}_4)}{\langle 12 \rangle \langle 14 \rangle}\right) \nn \\
& = \left( \frac{1}{\epsilon^2} S^{(0)} + \frac{1}{\epsilon} S^{(1)} \right) \mathcal{A}_4 \,,
\end{align}
if we expand the second line of above equations in terms of $\epsilon$ up to leading and sub-leading orders.

\subsection{Soft limit of the 8-point amplitude}

In this last subsection, we examine the soft limit of the 8-point amplitude we computed in section~\ref{sec:8pt}. In view of the computational complexity, 
we content ourselves with checking the leading order soft factor $S^{(0)}$.

\subsubsection*{8-point amplitude with u-cyclic gauge}  

To take the double soft limit of the 8-point amplitude, we revisit the computation of section~\ref{sec:8pt} with two slight changes. 
The first is that, to be specific, we work in the u-cyclic gauge. The $C$-matrix 
is given by 
\begin{align} 
C = 
\begin{pmatrix}
1 & 1& -\hat{u}_{12}   &  \hat{u}_{12}   & -\hat{u}_{13}   & \hat{u}_{13}   & - \hat{u}_{14}   &  \hat{u}_{14}    \\
\hat{u}_{12}   & -\hat{u}_{12}   & 1 & 1 & -\hat{u}_{23}   &  \hat{u}_{23}   & - \hat{u}_{24}   & \hat{u}_{24}   \\
\hat{u}_{13}   & -\hat{u}_{13}   & \hat{u}_{23}   & -\hat{u}_{23}   & 1 & 1 & -\hat{u}_{34}   &  \hat{u}_{34}  \\
\hat{u}_{14}   & -\hat{u}_{14}   & \hat{u}_{24}   & -\hat{u}_{24}   & \hat{u}_{34}  & -\hat{u}_{34}  & 1 & 1
\end{pmatrix}\,, 
\quad \hat{u}_{mn} = \hat{u}_{mn}(z) \,.
\end{align}
In this gauge, the fermionic delta function reduces to
\begin{align}
\delta^{12}(C \cdot \eta)
& = \prod_{I=1}^{3} \frac{1}{4!} \epsilon^{mnpq}
(\bar{\theta}_m^I+\hat{u}_{mi}(z) \theta^{iI}) (\bar{\theta}_n^I+\hat{u}_{nj}(z) \theta^{jI}) (\bar{\theta}_p^I+\hat{u}_{pk}(z) \theta^{kI})( \bar{\theta}_q^I+\hat{u}_{ql}(z) \theta^{lI}) \nn \\
& = J_8^F \delta^{6}(Q) \prod_{I=1}^{3}(A_I z^2 + B_I z + C_I) 
\,, \quad J_8^F = \left(\frac{4}{R}\right)^3 \,.
\end{align}
The fermionic bilinear coefficients are 
\begin{align}
A_I  =  - \frac{1}{4} \epsilon_{mnpq} \bar{u}^{mn}_* \theta^{pI} \theta^{qI} \,, \quad 
B_I  =  \frac{1}{4} \epsilon^{mnpq} \epsilon_{prxy}\bar{u}^*_{mn} \bar{u}_*^{xy} \theta^{rI} \bar{\theta}_q^I  \,, 
\quad 
C_I  =\frac{1}{4} \epsilon^{mnpq} u_{mn}^* \bar{\theta}_p^I \bar{\theta}_q^I \,. 
\end{align}
The second, more important change compared to section~\ref{sec:8pt} is that, 
in order to expose the soft limit more clearly (more on this below), we use the contours $\mathcal{C}_2$ and $\mathcal{C}_4$ instead of $\mathcal{C}_1$ and $\mathcal{C}_3$. Of course the two choices are equal up to an overall sign.
In the notations of section~\ref{sec:8pt}, the result is 
\begin{align}
\mathcal{A}_{8}= \delta^3(P) \delta^6(Q) (1+\pi) J_8^B J_8^F  \left(\frac{F(2)}{\Delta_{21} \Delta_{23} \Delta_{24}}+\frac{F(4)}{\Delta_{41} \Delta_{42} \Delta_{43}} \right) \,, \quad J_8^B  = \frac{1}{2R} \,.
\end{align}
The $\lambda$-parity operator $\pi$ acts on $\mathcal{A}_8$ as 
\begin{align}
\pi : \Lambda_{8} \rightarrow - \Lambda_{8} \,.
\end{align}
The numerators $F(2)$ and $F(4)$ are given by
\begin{align}
F(2) &= - K_{22} J_{21}{}^{(5} J_{23}{}^{6} J_{24}{}^{7)} + \frac{3}{4} L^2{}_{(1}{}^2{}_3 J_{4)2(5} L^2{}_{6}{}^2{}_{7)} \,,  \\
F(4) &= - K_{44} J_{41}{}^{(5} J_{43}{}^{6} J_{42}{}^{7)} + \frac{3}{4} L^4{}_{(1}{}^4{}_3 J_{2)4(5} L^4{}_{6}{}^4{}_{7)} \,,
\end{align} 
The physical poles are related to the $\Delta_{ij}$ factors through
\begin{align}
\Delta_{21} &= - \frac{2^{10}}{R^4} p_{234}^2 \; p_{678}^2 \,, 
\quad \Delta_{23} = - \frac{2^{10}}{R^4} p_{345}^2 \; p_{781}^2 \,, 
\nn \\
\Delta_{41} &= - \frac{2^{10}}{R^4} p_{123}^2 \; p_{567}^2 \,, 
\quad \Delta_{43} = - \frac{2^{10}}{R^4}p_{456}^2 \; p_{812}^2 \,.
\end{align} 
The derivation of these relations is essentially the same as the one given in appendix \ref{sec:NN}. 
The factor $\Delta_{24}$ corresponds to spurious poles. 

When we consider the limit in which particles $7$ and $8$ become soft, divergent terms come from $\Delta_{21}$ and $\Delta_{23}$. If we use the contours $\mathcal{C}_1$ and $\mathcal{C}_3$ as in section~\ref{sec:8pt}, 
the two contributions are divided into two different on-shell diagrams. 
But, if we use the contours $\mathcal{C}_2$ and $\mathcal{C}_4$, 
both contributions come from the residues of $M_2(z)$ and we can ignore 
the residues of $M_4(z)$. 

\subsubsection*{Soft limit of 8-point amplitude} 

In the lightcone coordinates, the the double soft limit of the $7$ and $8$ 
is realized by
\begin{align}
w_{4} \rightarrow \epsilon \, w_{4} \,, \quad 
v^{4} \rightarrow \epsilon \, v^{4} \,.
\end{align}
In the $\epsilon \rightarrow 0$ limit, $u_{m4}^*$ and $\bar{u}^{n4}_*$ are of order $\epsilon$. As we discussed earlier, kinematic invariants receive $\epsilon^2$ corrections, so we can freely use the kinematic relations of 
the 6-point amplitude. For example, the identity \eqref{uubar} in the soft limit implies that 
\begin{align}
1 + u_{12}^*\bar{u}^{12}_* + u_{23}^*\bar{u}^{23}_* + u_{13}^*\bar{u}^{13}_*=0 +\mathcal{O}(\epsilon^2) \,.
\end{align}
To the leading order in $\epsilon$, the coefficients of minors $M_i(z)$  are given by
\begin{align}
a_{1} & = 4 \epsilon \bar{u}^{12}_* \bar{u}^{34}_* \,, \quad 
b_{1} = 4(\bar{u}^{12}_* u_{12}^* +1)  \,, \quad
c_{1} = 4\epsilon u_{12}^* u_{34}^*   \,, 
\nn \\
a_{2} & = 2\epsilon  (-\bar{u}^{24}_*+\bar{u}^{23}_*\bar{u}^{34}_*+\bar{u}^{12}_*\bar{u}^{14}_* )  \,, \quad
b_{2}  = 2(u_{12}^{*}\bar{u}^{23}_*+u_{23}^{*}\bar{u}^{12}_* + u_{13}^{*} + \bar{u}^{13}_*)   \,, \quad
\nn \\
& \qquad 
c_{2}  = 2\epsilon (-u_{24}^{*}+u_{23}^{*}u_{34}^{*}+u_{12}^{*}u_{14}^{*})  \,, \nn \\
a_{3} & = 4\epsilon \bar{u}^{14}_* \bar{u}^{23}_*    \,, \quad
b_{3}  = 4(u_{23}^{*}\bar{u}^{23}_* +1)  \,, \quad
c_{3}  = 4\epsilon u_{14}^{*}u_{23}^{*}  \,, 
\nn \\ 
a_{4} & = 2(- \bar{u}^{13}_*+\bar{u}^{12}_*\bar{u}^{23}_*) \,, \quad 
b_{4} = 2\epsilon (u_{12}^{*}\bar{u}^{14}_*+u_{23}^{*}\bar{u}^{34}_*+u_{34}^{*}\bar{u}^{23}_*+u_{14}^{*}\bar{u}^{12}_* + u_{24}^{*} + \bar{u}^{24}_* ) \,,
\nn \\
&\qquad
c_{4} = 2(-u_{13}^{*}+u_{12}^{*}u_{23}^{*})   \,.
\end{align}

If we focus on the leading order only, 
the supermomentum-conserving delta function of $\mathcal{A}_8$ trivially reduces to  that of $\mathcal{A}_6$: 
\begin{align}
\delta^6 (Q_8)|_{\epsilon \rightarrow 0} = \delta^6 (Q_6) \,.
\end{align} 
The only non-vanshing contribution from the fermionic part in the numerator $F(2)$ is 
\begin{align}
F(2) |_{\epsilon \rightarrow 0} 
= -(b_2)^3  c_4 \alpha_{12} \alpha_{23} \; a_5 a_6 a_7 + (b_2)^3 a_4 \gamma_{12} \gamma_{23} \; c_5 c_6 c_7  \,.
\end{align}
One can easily check that
fermionic bilinears $a_{4+I}=A_I$ and $c_{4+I}=C_I$  become
\begin{align}
A_I |_{\epsilon \rightarrow 0} & = - \frac{1}{2} \epsilon_{mnp} \bar{u}^{mn}_* \theta^{pI} \theta^{4I}  = - \zeta_-^I \times \theta^{4I} \,,\\
C_I |_{\epsilon \rightarrow 0} & = \frac{1}{2} \epsilon^{mnp} u_{mn}^* \bar{\theta}_p^I \bar{\theta}_4^I  = \zeta_+^I \times \bar{\theta}_4^I \,.
\end{align}
The soft limit of $\Delta_{ij}$ gives
\begin{align}
\Delta_{21} = -\alpha_{12} \gamma_{12} \,, \quad
\Delta_{23} = -\alpha_{23} \gamma_{23} \,, \quad
\Delta_{24} = (b_2)^2 a_4 c_4 \,.
\end{align}
We observe that the following useful identities hold in the soft limit:
\begin{align}
a_4 &= - M_1^- \,, \\
c_4 &= - M_1^+ \,, \\
\alpha_{12} & = -8\epsilon \frac{\langle 67 \rangle + \langle 68 \rangle}{R} \times M_2^+ 
= -8\epsilon \beta_+ \frac{\langle 16 \rangle }{R} M_2^+ \,, 
\\
\gamma_{12} & = - 8\epsilon \frac{\langle 67 \rangle - \langle 68 \rangle}{R} \times M_2^-
=  - 8\epsilon \beta_- \frac{\langle 16 \rangle }{R} M_2^- \,, 
\\
\alpha_{23} & = 8\epsilon \frac{\langle 17 \rangle + \langle 18 \rangle}{R} \times M_3^+ 
= 8\epsilon \alpha_+ \frac{\langle 16 \rangle }{R} M_3^+ \,, 
\\
\gamma_{23} & = 8\epsilon \frac{\langle 17 \rangle -  \langle 18 \rangle}{R} \times  M_3^- 
= 8\epsilon \alpha_- \frac{\langle 16 \rangle}{R} M_3^- \,, 
\end{align}
where $M_{i}^{\pm}$ here denote consecutive minors of $C^\pm$ contributing to $\mathcal{A}_6$.

Collecting all ingredients, we obtain the soft limit of the 8-point amplitude 
in the leading order 
\begin{align}
\mathcal{A}_{8}|_{\epsilon \rightarrow 0}
& = (1+\pi)  \frac{2^5}{ R^4} \frac{F(2)}{\Delta_{21} \Delta_{23} \Delta_{24}} \nn \\
& = (1+\pi) \left( \frac{2^5}{ R^4} \right)
\frac{ -(b_2)^3 c_4 \alpha_{12} \alpha_{23} \; a_5 a_6 a_7 + (b_2)^3 a_4 \gamma_{12} \gamma_{23} \; c_5 c_6 c_7 }
{(b_2)^2 a_4 c_4 \alpha_{12} \alpha_{23} \gamma_{12} \gamma_{23}} 
\nn \\
& = (1+\pi) \left( \frac{2^5 b_2}{ R^4} \right)
\left(  -\frac{1}{ a_4 \gamma_{12} \gamma_{23}} a_5 a_6 a_7 + \frac{1}{c_4 \alpha_{12} \alpha_{23}} c_5 c_6 c_7 \right) 
\nn \\
& =  (1+\pi) \left(  \frac{ b_2 R }{ 32 \langle 16 \rangle} \right) 
\left[ 
\left( \frac{\delta^3(\theta_4)}{2 \epsilon^2 \langle 16 \rangle \alpha_-\beta_-} \right)
\frac{ 32 \delta^3(\zeta_-)}{ R^3 M_1^- M_2^- M_3^-} 
+ 
\left\{ 
(+)\leftrightarrow (-) , 
\theta \leftrightarrow \bar{\theta}
\right\}
\right] 
\nn \\
& = 
\frac{1}{\epsilon^2} S^{(0)} \mathcal{A}_6 + \mathcal{O}(1/\epsilon) \,.
\end{align}
In the final step, we used the following non-trivial identity
\begin{align}
\frac{b_2 R }{\langle 16 \rangle} 
& = \frac{2R}{\langle 16 \rangle} \left( u_{12}^{*}\bar{u}^{23}_*+u_{23}^{*}\bar{u}^{12}_* + u_{13}^{*} + \bar{u}^{13}_* \right) \nn \\
& = \frac{2}{\langle 16 \rangle} \left( \langle w_1 v^3 \rangle - \langle v^1 w_3 \rangle + \langle w_1 w_3 \rangle - \langle v^1 v^3 \rangle \right) 
 = 8 \,.
\end{align}



\vskip 1cm 
\acknowledgments

%
The work of SL and YY is supported by Samsung Science and Technology Foundation under Project Number SSTF-BA1402-08.

\vskip 2cm

\newpage 

\appendix
\centerline{\Large\bf Appendix}


\section{Closer look at factorization channels}

\subsection{6-point amplitude \label{sec:MM}}

From the Grassmannian integral point of view, 
the three factorization channels of the 6-point amplitude 
originate from the product of minors from the two branches, 
when the minors are computed from the $C$ matrix satisfying 
$C\cdot\lambda=0$. 
In the alternating signature metric $g=\mathrm{diag}(-+-+-+)$, 
the precise relation can be written as 
\begin{align}
M^+_{ijk} M^-_{ijk} = K (-1)^{i+j+k} p_{ijk}^2 \,.
\label{MM-ijk}
\end{align}
The subscripts $(ijk)$ denote a set of three columns from the $C$-matrix,
and the superscript $(\pm)$ distinguishes the two branches. 
The prefactor $K$ may depend on the gauge choice but is independent 
of the choice of columns $(ijk)$. 

Since the relation \eqref{MM-ijk} is fully gauge-covariant, it is sufficient to prove it in any convenient gauge. Moving to another gauge alters the prefactor $K$, but the relation remains intact. We find it convenient to begin with a slightly unfamiliar ``$\lambda$-gauge":
\begin{align}
C^{\pm}_\lambda = 
\begin{pmatrix}
-\lambda_1^\alpha & \lambda_2^\alpha & -\lambda_3^\alpha & 
\lambda_4^\alpha & -\lambda_5^\alpha & \lambda_6^\alpha 
\\
\pm \langle 35 \rangle &  \langle 46 \rangle &
\pm \langle 51 \rangle &  \langle 62 \rangle &
\pm \langle 13 \rangle &  \langle 24 \rangle
\end{pmatrix} 
\,.
\label{lambda-gauge}
\end{align}
This gauge was used in deriving a dual Grassmannian formula for ABJM amplitudes \cite{Elvang:2014fja}. 

We can verify \eqref{MM-ijk} by explicit computations. Two well-known facts reduces the number of independent terms considerably. 
First, $(ijk)$ and its complement $(\overline{ijk})$ have minors that are equal up to an overall factor \cite{Huang:2013owa,Kim:2014hva}. In the $(k,k)$ signature, they satisfy 
\begin{align}
M^+_{ijk} M^-_{ijk} = - M^+_{\overline{ijk}} M^-_{\overline{ijk}}\,.
\end{align}
Second, the gauge \eqref{lambda-gauge} respects a permutation symmetry among odd labels and another permutation symmetry among even labels. So, without loss of generality, we may only check $(ijk)=(246)$ and $(ijk)=(456)$. 

The first one trivially gives
\begin{align}
M^+_{246} M^-_{246} = (\langle 24 \rangle^2 + \langle 46 \rangle^2 + \langle 62 \rangle^2)^2 = (p_{246}^2)^2 \,. 
\end{align}
The second one is slightly more involved. 
\begin{align}
M^+_{456} M^-_{456} &= ( \langle 24 \rangle  \langle 45 \rangle + \langle 56 \rangle  \langle 62 \rangle)^2 - \langle 46 \rangle^2  \langle 13 \rangle^2
\nn \\
&=\langle 24 \rangle^2  \langle 45 \rangle^2 + \langle 56 \rangle^2  \langle 62 \rangle^2 + 2 \langle 24 \rangle  \langle 45 \rangle\langle 56 \rangle  \langle 62 \rangle
\nn \\
&\qquad - \langle 46 \rangle^2 ( \langle 24 \rangle^2 + \langle 46 \rangle^2  +\langle 62 \rangle^2  - \langle 25 \rangle^2  - \langle 45 \rangle^2  - \langle 56 \rangle^2  )
\nn \\
&= ( \langle 24 \rangle^2 + \langle 46 \rangle^2 +\langle 62 \rangle^2) 
( \langle 45 \rangle^2 + \langle 56 \rangle^2 - \langle 46 \rangle^2 ) 
\nn \\
&\qquad +  \langle 25 \rangle^2  \langle 46 \rangle^2 
- (\langle 24 \rangle  \langle 56 \rangle - \langle 62 \rangle  \langle 45 \rangle )^2 
\nn \\
&= - p_{246}^2 p_{456}^2 \,.
\end{align}
In the second step, we used momentum conservation. Note that the alternating signature metric is reflected in the square of a partial sum of momenta as 
\begin{align}
p_{ijk}^2 = (-1)^{i+j} \langle ij \rangle^2 + (-1)^{j+k} \langle jk \rangle^2 + (-1)^{k+i} \langle ki \rangle^2\,.
\end{align}
In the last step, we used Schouten identity to cancel the last two terms.
To sum up, in the gauge \eqref{lambda-gauge}, we have proved the relation 
\eqref{MM-ijk} with $K_\lambda=p_{246}^2$.

Let us compute the gauge dependent factor $K$ in \eqref{MM-ijk} for the u-cyclic gauge used in the main text. 
Other u-type gauges can be treated similarly. 
There are two ways to compute $K$.  
We may either compute it directly in a u-gauge or find a gauge transformation 
between a u-gauge and the $\lambda$-gauge \eqref{lambda-gauge}. In the latter approach, we look for $GL(3,\mathbb{C})$ matrices $G^\pm$ satisfying
\begin{align}
C^\pm_\lambda = G^{\pm} C^\pm_u \,, 
\end{align}
where we recall the $C$-matrices in the u-cyclic gauge, 
\begin{align}
C_u^+ &= 
\begin{pmatrix}
1 & 1 & -u_{12} & u_{12} & -u_{13} & u_{13} \\
u_{12} & -u_{12} & 1 & 1 & -u_{23} & u_{23} \\
u_{13} & -u_{13} & u_{23} & -u_{23} & 1 & 1
\end{pmatrix}_{u = u^*} \,, 
\nn \\
C_u^- &= 
\begin{pmatrix}
-1 & 1 & \bar{u}^{12} & \bar{u}^{12} & \bar{u}^{13} & \bar{u}^{13} \\
-\bar{u}^{12} & -\bar{u}^{12} & -1 & 1 & \bar{u}^{23} & \bar{u}^{23} \\
-\bar{u}^{13} & -\bar{u}^{13} & -\bar{u}^{23} & -\bar{u}^{23} & -1 & 1
\end{pmatrix}_{\bar{u} = \bar{u}_*} \,.
\end{align}
The matrices $G^\pm$ turn out to take a simple form
\begin{align}
G^\pm = \frac{1}{2} 
\begin{pmatrix}
\mp \lambda_1^\alpha + \lambda_2^\alpha &  
\mp \lambda_3^\alpha + \lambda_4^\alpha &  
\mp \lambda_5^\alpha + \lambda_6^\alpha 
\\
\langle 35 \rangle + \langle 46 \rangle & 
\langle 51 \rangle + \langle 62 \rangle & 
\langle 13 \rangle + \langle 24 \rangle 
\end{pmatrix} \,.
\end{align}
To make contact with the u-cyclic gauge, 
we note that the first two rows of $G^\pm$ are light-cone coordinates 
and the third row can be rewritten as 
\begin{align}
(\langle 35 \rangle + \langle 46 \rangle ,
\langle 51 \rangle + \langle 62 \rangle ,
\langle 13 \rangle + \langle 24 \rangle ) = 
\frac{R}{2} (u^*_{23}-\bar{u}_*^{23}, u^*_{31}-\bar{u}_*^{31}, u^*_{12}-\bar{u}_*^{12}) \,.
\end{align}
Using this and the identity \eqref{uubar}, we find $\det(G^\pm)$, 
\begin{align}
\det(G^+) &= \left(R/4\right)^2 \left[-(u^*_{23}-\bar{u}_*^{23})\bar{u}_*^{23}-(u^*_{31}-\bar{u}_*^{31})\bar{u}_*^{31}- (u^*_{12}-\bar{u}_*^{12})\bar{u}_*^{12}\right] 
\nn \\
&=\left(R/4\right)^2  \left[ 1+ (\bar{u}_*^{12})^2 + (\bar{u}_*^{23})^2 +(\bar{u}_*^{31})^2 \right] = \left(R/4\right)^2  (M^-_{246})_{u} \,,
\nn \\
\det(G^-) &= \left(R/4\right)^2 \left[(u^*_{23}-\bar{u}_*^{23})u^*_{23}+(u^*_{31}-\bar{u}_*^{31})u^*_{31}+(u^*_{12}-\bar{u}_*^{12})u^*_{12}\right] 
\nn \\
&=\left(R/4\right)^2  \left[ 1+ (u^*_{12})^2 + (u^*_{23})^2 +(u^*_{31})^2 \right] = \left(R/4\right)^2  (M^+_{246})_{u} \,.
\end{align}
The determinants allow us to determine $K_u$ up to a sign, 
\begin{align}
p_{246}^2 = K_\lambda = \det(G^+)\det(G^-)K_u = (R/4)^4(M^-_{246} M^+_{246})_{u} 
=  (R/4)^4 K_u^2 (p_{246}^2) \,.
\end{align}
To fix the sign and make a cross-check, we compute $M^+M^-$ directly in the u-cyclic gauge. We begin by rewriting $p_{246}^2$ in the light-cone variables, 
\begin{align}
4^2 p_{246}^2 &= 4^2(\langle 24 \rangle^2  + \langle 46 \rangle^2 + \langle 62 \rangle^2 )
\nonumber \\
&= \langle v_1+w_1,v_2+w_2 \rangle^2 + \langle v_2+w_2, v_3+w_3 \rangle^2 +\langle v_3+w_3, v_1+w_1 \rangle^2  \nonumber \\
&= (v_{12} +w_{12}+ \langle w_1v_2 \rangle + \langle v_1w_2\rangle )^2 + \mbox{(cyclic)}\,. 
\end{align}
We introduced the short-hand notations, $v_{mn} = \langle v^m v^n \rangle$, 
$w_{mn} = \langle w_m w_n \rangle$. 
Using the identity $R\langle v^m w_n \rangle = - v^{mp}w_{np}$, the above identity can be further rewritten as
\begin{align}
(4R)^2 p_{246}^2&= (R v_{12} + R w_{12} + v_{23}w_{13}-v_{13}w_{23}  )^2 
+ \mbox{(cyclic)}  
\nonumber \\
&= R^2 (v_{12}+w_{12})^2 +(v_{23}w_{13}-v_{13}w_{23})^2  
+ \mbox{(cyclic)} 
\nonumber \\
&=R^2(v_{12}^2+v_{13}^2+v_{23}^2) +R^2(w_{12}^2+w_{13}^2+w_{23}^2) +2 R^4 \nonumber \\
& \qquad +(v_{23}w_{13}-v_{13}w_{23})^2+(v_{32}w_{12}-v_{12}w_{32})^2 +(v_{31}w_{21}-v_{21}w_{31})^2  
\nonumber \\
&=(R^2+v_{12}^2+v_{13}^2+v_{23}^2)( R^2 +w_{12}^2+w_{13}^2+w_{23}^2) 
\nn \\
&\qquad + R^4- (v_{12}w_{12}+v_{13}w_{13} +v_{23}w_{23})^2 
\nonumber \\
&=(R^2+v_{12}^2+v_{13}^2+v_{23}^2)( R^2 +w_{12}^2+w_{13}^2+w_{23}^2) 
= R^4 M_{246}^- M_{246}^+ \,.
\end{align}
In the second line, all cross terms cancelled out when summed over cyclic permutation. In the third and fifth line, we used  
$R^2 = v_{12} w_{12} +  v_{23} w_{23} + v_{31} w_{31}$, 
which is the same as \eqref{uubar}. 
To conclude, we have verified \eqref{MM-ijk} that in the u-cyclic gauge, 
\begin{align}
M^+_{ijk} M^-_{ijk} = \left(\frac{4}{R} \right)^2 (-1)^{i+j+k} p_{ijk}^2 \,.
\label{MM-ijk-u}
\end{align}


\subsection{8-point amplitude \label{sec:NN}}

In this section, we work in the u-factorization gauge introduced in section \ref{sec:u-gauge}:
\begin{align}
C=
\begin{pmatrix} 
1 & \hat{u}_{12}& \hat{u}_{13} & \hat{u}_{14} & 1 & -\hat{u}_{12} & -\hat{u}_{13} & -\hat{u}_{14} 
\\
-\hat{u}_{12} & 1 & \hat{u}_{23} & \hat{u}_{24} & \hat{u}_{12}& 1 & -\hat{u}_{23} & -\hat{u}_{24}
\\
-\hat{u}_{13} & -\hat{u}_{23} & 1 & \hat{u}_{34}& \hat{u}_{13} & \hat{u}_{23} & 1 & -\hat{u}_{34} 
\\
-\hat{u}_{14} & -\hat{u}_{24} & -\hat{u}_{34} & 1 & \hat{u}_{14} & \hat{u}_{24} & \hat{u}_{34} & 1 
\end{pmatrix}
\,, 
\quad \hat{u}_{mn} = \hat{u}_{mn}(z) \,.
\end{align}
As we mentioned earlier, the minors are quadratic polynomials in $z$. For later convenience, we organize the coefficients of the polynomials with some extra shorthand notations, 
\begin{align}
M_1&=z^2 + (\hat{u}_{12}^2 +\hat{u}_{13}^2+\hat{u}_{14}^2+\hat{u}_{23}^2+\hat{u}_{24}^2+\hat{u}_{34}^2 ) +1 
\nonumber \\
&= (1+V)z^2 + 2 U z + (1+W) \,,
\nn \\
M_2 &= z^2 - (\hat{u}_{23}^2+\hat{u}_{34}^2+\hat{u}_{24}^2)+(\hat{u}_{12}^2+\hat{u}_{13}^2+\hat{u}_{14}^2  ) -1 \nonumber \\
&=(1 + r_2 - s_2 )z^2 -2(p_2-q_2)z  -(1+ m_2-n_2 ) \,,
\nn \\
M_3 &=z^2 +(\hat{u}_{12}^2+\hat{u}_{34}^2)-(\hat{u}_{13}^2 +\hat{u}_{24}^2+\hat{u}_{14}^2+\hat{u}_{23}^2  )  +1 \nonumber \\
&=(1 + r_{3}-s_3 )z^2 +2(p_3-q_3)z +(1+ m_3-n_3 ) \,,
\nn \\
M_4 &=z^2 -(\hat{u}_{12}^2+\hat{u}_{13}^2+\hat{u}_{23}^2)+(\hat{u}_{14}^2+\hat{u}_{24}^2+\hat{u}_{34}^2  )  -1 \nonumber \\
&=(1 + r_4-s_4 )z^2 -2(p_4-q_4)z -(1+ m_4-n_4 ) \,.
\end{align}
We introduced 
\begin{align}
V = \frac{1}{2} \sum_{i,j}(\bar{u}_{*}^{ij})^2\,,
\quad
U = \frac{1}{4} \sum_{i,j,k,l} \epsilon_{ijkl}u^*_{ij}\bar{u}_*^{kl} \,,
\quad 
W = \frac{1}{2} \sum_{i,j}(u^*_{ij})^2 \,,
\end{align} 
and 
\begin{align}
\begin{array}{rclcrcl}
r_2 &=& (\bar{u}_*^{23})^2+(\bar{u}_*^{34})^2+(\bar{u}_*^{24})^2\,, & 
&s_2 &=& (\bar{u}_*^{12})^2+(\bar{u}_*^{13})^2 +(\bar{u}_*^{14})^2\,, 
\\
p_2 &=& u^*_{23}\bar{u}_*^{14}-u^*_{24}\bar{u}_*^{13}+u^*_{34}\bar{u}_*^{12}\,, & 
&q_2 &=& u^*_{12}\bar{u}_*^{34}-u^*_{13}\bar{u}_*^{24} +u^*_{14}\bar{u}_*^{23}\,, 
\\
m_2 &=& (u^*_{23})^2+(u^*_{34})^2 +(u^*_{24})^2 \,, & 
&n_2 &=& (u^*_{12})^2+(u^*_{13})^2 +(u^*_{14})^2 \,,
\\
r_3 &=& (\bar{u}_*^{12})^2+(\bar{u}_*^{34})^2 \,, &
&s_3 &=& (\bar{u}_*^{13})^2+(\bar{u}_*^{24})^2+(\bar{u}_*^{14})^2+(\bar{u}_*^{23})^2, 
\\
p_3 &=& u^*_{12}\bar{u}_*^{34} +u^*_{34}\bar{u}_*^{12} \,, &
&q_3 &=& u^*_{14}\bar{u}_*^{23} +u^*_{23}\bar{u}_*^{14}-u^*_{13}\bar{u}_*^{24}-u^*_{24}\bar{u}_*^{13} \,, 
\\
m_3 &=& (u^*_{12})^2+(u^*_{34})^2 \,, & 
&n_3 &=& (u^*_{13})^2+(u^*_{14})^2 +(u^*_{23})^2+(u^*_{24})^2 \,,
\\
r_4 &=& (\bar{u}_*^{12})^2+(\bar{u}_*^{13})^2+(\bar{u}_*^{23})^2 \,,  &
&s_4 &=& (\bar{u}_*^{14})^2+(\bar{u}_*^{24})^2+(\bar{u}_*^{34})^2 \,,  
\\
p_4 &=& u^*_{12}\bar{u}_*^{34}-u^*_{13}\bar{u}_*^{24}+u^*_{23}\bar{u}_*^{14}\,, & 
&q_4 &=& u^*_{14}\bar{u}_*^{23}-u^*_{24}\bar{u}_*^{13}+u^*_{34}\bar{u}_*^{12}\,, 
\\
m_4 &=& (u^*_{12})^2+(u^*_{13})^2 +(u^*_{23})^2 \,, &
\qquad 
&n_4 &=& (u^*_{14})^2+(u^*_{24})^2 +(u^*_{34})^2 \,.
\end{array}
\end{align}
Note that $r_h+s_h  = V$, $p_h+q_h = U$, $m_h+n_h = W$ for each $h=2,3,4$.
The quantities $(\alpha_{ij},\beta_{ij},\gamma_{ij})$ can be expressed in terms of $(m_h,n_h,p_h,q_h,r_h,s_h)$ in a simple way:
\begin{align}
\alpha_{12} &= -4\left[ (m_2+1)q_2 - n_2 p_2 \right] \,, \quad 
\gamma_{12}  = -4\left[ p_2 (r_2+1) - q_2 s_2  \right] \,, \nn \\
\beta_{12} &= 2\left[ (m_2+1)(r_2+1) - n_2 s_2 \right] \,, \nn \\
\alpha_{13} &= 4\left[ (m_3+1)q_3 - n_3 p_3 \right] \,, \quad
\gamma_{13}  = 4\left[  p_3 s_3 - q_3 (r_3+1)  \right] \,, \nn \\
\beta_{13} &= -2\left[ (m_3+1)s_3 - n_3 (r_3+1) \right] \,, \nn \\
\alpha_{14} &=-4\left[ (m_4+1)q_4 - n_4 p_4 \right] \,, \quad 
\gamma_{14}  =-4\left[ p_4 (r_4+1) - q_4 s_4  \right] \,, \nn \\
\beta_{14} &=  2\left[ (m_4+1)(r_4+1) - n_4 s_4 \right] \,. 
\label{abc14}
\end{align}

As explained in the main text, the two physical poles for $p_{123}^2$ and $p_{567}^2$ are captured by $\Delta_{14}$. We can verify this by an explicit computation. With a slight rewriting to facilitate comparison with the u-gauge, we have
\begin{align}
\left(\frac{4}{R}\right)^2 p_{123}^2 
&=\left(\frac{4}{R}\right)^2 (\langle 12 \rangle^2 + \langle 13 \rangle^2 + \langle 23 \rangle^2) 
\nonumber \\
&=(1+r_4)(1+m_4)-2(\bar{u}_*^{14}u^*_{14}+\bar{u}_*^{24}u^*_{24}+\bar{u}_*^{34}u^*_{34})^2 +s_4n_4 
\nonumber \\
&\quad + \left( (\bar{u}_*^{12}+u^*_{12}) +2(\bar{u}_*^{23}u^*_{13}-\bar{u}_*^{13}u^*_{23}) \right)(\bar{u}_*^{24}u^*_{14}-\bar{u}_*^{14}u^*_{24}) \nonumber \\
&\quad +\left( (\bar{u}_*^{13}+u^*_{13}) +2(\bar{u}_*^{32}u^*_{12}-\bar{u}_*^{12}u^*_{32}) \right)(\bar{u}_*^{34}u^*_{14}-\bar{u}_*^{14}u^*_{34}) \nonumber \\
&\quad + \left( (\bar{u}_*^{23}+u^*_{23}) +2(\bar{u}_*^{31}u^*_{21}-\bar{u}_*^{21}u^*_{31}) \right)(\bar{u}_*^{34}u^*_{24}-\bar{u}_*^{24}u^*_{34})  
\nn \\
&\equiv (1+r_4)(1+m_4)-2(t_4)^2 +s_4n_4 
\nn \\
&\quad + (x_{12} + 2 y_{12}) z_{12}  
+ (x_{23} + 2 y_{23}) z_{23}  
+ (x_{31} + 2 y_{31}) z_{31}  \,,
\end{align}
and, similarly, 
\begin{align}
\left(\frac{4}{R}\right)^2 p_{567}^2 
&=
\left(\frac{4}{R}\right)^2(\langle 56 \rangle^2 + \langle 57 \rangle^2 +\langle 67 \rangle^2) 
\nonumber \\
&= (1+r_4)(1+m_4)-2(t_4)^2 +s_4n_4 
\nn \\
&\quad + (-x_{12} + 2 y_{12}) z_{12}  
+ (-x_{23} + 2 y_{23}) z_{23}  
+ (-x_{31} + 2 y_{31}) z_{31}  \,.
\end{align}
Combining the two, we have
\begin{align}
\frac{2^8}{R^4}p_{123}^2p_{567}^2 
&=
\left[ (1+r_4)(1+m_4)-2(t_4)^2 +s_4n_4 + 2 ( y_{12} z_{12} + y_{23} z_{23} +y_{31} z_{31}) \right]^2 
\nn \\ 
&\quad - ( x_{12} z_{12} + x_{23} z_{23} + x_{31} z_{31})^2 \,.
\label{p123p567}
\end{align}
On the other hand, we deduce from \eqref{abc14} that 
\begin{align}
\frac{1}{4}\Delta_{14}&= \left[(m_4 +1)(r_4+1) - n_4 s_4 \right]^2 -4\left[(m_4+1)q_4 - n_4 p_4 \right] \left[p_4(r_4+ 1) -q_4 s_4  \right] \,.
\label{delta14}
\end{align}
A lengthy but straightforward computation shows that \eqref{p123p567} and \eqref{delta14} are equal.

\end{document}